\documentclass[12pt,preprint]{aastex}

\newcommand{\dgr}{$^{\circ}$}
\newcommand{\etal}{et~al.\ }
\newcommand{\kms}{km~s$^{-1}$}

\newcommand{\HB}{H$\beta\/$}
\newcommand{\HC}{H$\gamma\/$}

\usepackage[twocolumn]{emulateapj5}

\slugcomment{To appear in The Astrophysical Journal}

\shorttitle{Spectropolarimetry of 2MASS QSOs}
\shortauthors{Smith \etal}

\begin{document}

\title{Optical Spectropolarimetry of Quasi-Stellar Objects Discovered
by the Two-Micron All Sky Survey{\footnotemark}}
\footnotetext{A portion of the results presented here made use
of the Multiple Mirror Telescope Observatory, a facility operated
jointly by the University of Arizona and the Smithsonian Institution.}

\author{Paul S. Smith, Gary D. Schmidt, and Dean C. Hines}
\affil{Steward Observatory, The University of Arizona,
    Tucson, AZ 85721}

\and

\author{Craig B. Foltz}
\affil{Multiple Mirror Telescope Observatory, 
The University of Arizona, Tucson, AZ 85721}

\begin{abstract}

Highly polarized QSOs discovered in the Two-Micron All Sky Survey (2MASS)
have been observed to determine the source(s) of optical polarization
in this near-infrared color-selected sample.
Broad emission lines are observed
in the polarized flux spectra of most objects, and the polarization 
of the lines is at about the same level and position angle as the
continuum.
Generally, the 
continuum is bluer and the broad-line Balmer decrement is smaller
in polarized light than for the spectrum of total flux.
Narrow emission lines are much less polarized than the broad lines and
continuum for all polarized objects.
These properties favor scattering by material close to
a partially obscured and reddened active
nucleus,
but exterior to the regions producing the broad-line emission,
as the source of polarized flux in 2MASS QSOs.
The largely unpolarized narrow-line features require that the electrons
or dust polarizing the light be located at distances from the 
nucleus not much greater than the extent of the narrow emission-line
region.
The conclusion that the scattering material is located close to the 
nucleus is reinforced by the observation in four objects of changes 
in both the degree and position angle of polarization across the
broad H$\alpha\/$ emission-line profile,
indicating that the broad emission-line region (BLR)
is at least partially resolved at the distance of the scatterers.
In addition to known high-polarization objects, four 2MASS QSOs
with AGN spectral types of 1.9 and 2 were observed to search for 
hidden BLRs.
Broad lines were detected in polarized light for two of these objects, and
the polarizing mechanism appears to be the same for these objects as
for the highly polarized QSOs in the sample that readily show
broad emission lines in their spectra.
The small observed sample of eight Type~1 2MASS QSOs has weak
[\ion{O}{3}] emission in comparison to optically-selected AGN with
similar near-infrared luminosity.
The observations also show that starlight from the host galaxy
contributes a significant amount of optical flux, especially for
the narrow-line objects, and support the suggestion that many
2MASS QSOs are measured to have low polarization simply
because of dilution of the polarized AGN light by the host
galaxy.

\end{abstract}


\section{Introduction}

The Two-Micron All Sky Survey \citep[2MASS;][]{skrutskie97} has 
revealed previously unknown, primarily low-redshift
active galactic nuclei (AGN) whose space density likely exceeds
that of AGN selected by their ultraviolet and optical colors
\citep{cutri01}.
This large population of radio-quiet AGN was found by \citet{cutri01}
using a simple near-IR
color criterion ($J - K_s > 2$) with no regard for optical, radio, or
X-ray properties.
Secondary selection criteria are that the plane of the Galaxy
is avoided ($\vert b \vert > 30$\dgr\/) to
minimize the contamination of the AGN survey by reddened galactic objects,
and that inclusion in the AGN sample requires detection in all three 2MASS
near-IR bandpasses.
The latter requirement allows for a well-defined color-selected
sample, but at the expense of missing even redder AGN given the sensitivity
limits of 2MASS.
The color criterion ensures that known AGN (as well as stars) are
not a major contaminant in
the survey since the vast majority of cataloged AGN have bluer $J - K_s\/$
colors.

The simple selection criteria adopted by \citet{cutri01} result in 
an efficient survey for low-redshift ($z < 0.7$)
AGN that are missed in surveys
using traditional optical
and ultraviolet search methods.
Unlike UV-excess AGN samples,
the 2MASS objects encompass a large range of AGN optical
spectral types.
Broad emission-line (Type~1), narrow emission-line
(Type~2), 
and intermediate (Type 1.5--1.9) objects are all well-represented.
\citet{cutri01} find $\sim 3\times$ the number of
AGN showing broad emission
lines in their spectra than Type~2 AGN; over the entire sky this
fraction translates to nearly 6000 Type~2 AGN with $K_s \leq 15$.
In addition to the red near-IR colors of the 2MASS objects,
their optical faintness and red 
optical colors suggest that the survey is uncovering a
large population of dust-obscured AGN.

Follow up X-ray observations and optical broadband polarimetry
of 2MASS AGN with
QSO-like near-IR luminosities support the contention that the bulk
of the near-IR--selected sample is composed of objects
at least partially obscured from
our direct line of sight.
\citet{wilkes02} find large absorbing columns
($N_H \sim 10^{21}$--10$^{23}$~cm$^{-2}$) toward the nuclear X-ray
sources.
\citet{smith02} find that a large
fraction ($>$10\%) of luminous 2MASS AGN
($M_{K_s} \lesssim -25$) are highly polarized ($P > 3$\%)
compared to optically-selected QSOs
and broad absorption-line
QSOs.
In fact, broadband polarizations as high as $\sim 10$\% are measured for
a few of these near-IR--selected QSOs.

In this paper, optical spectropolarimetry of all highly polarized and some
moderately ($P = 1$--3\%)
polarized QSOs found by \citet{smith02} is presented in an effort to
better understand the
polarizing mechanism(s) and thus place constraints on the nature of
the 2MASS sample and AGN phenomena in general.
In particular, it is of interest to test if the polarization 
properties of near-IR--selected QSOs are consistent with those
found in other highly polarized AGN samples where it has been
shown that much of the observational data can be explained in
the terms of an obscuring dust torus surrounding the active
nucleus
\citep[e.g.,][]{antonucci93}.
For many Seyfert~2 galaxies
\citep{antonucci85,miller90,tran92},
narrow-line radio galaxies
\citep[NLRGs;][]{tranc95,ogle97,cohen99},
and 
hyperluminous infrared galaxies
\citep[HIGs; see e.g.,][]{wills92,young96,hines01},
spectropolarimetry has been able to show that orientation plays
a critical role 
in the classification of AGN.
The basic result for all of these narrow emission-line objects is that 
light from the active nucleus, including the ionizing continuum
and the emission from the broad-line region (BLR), is obscured
from direct view by dust near the nucleus, but for lines of sight 
that do not intersect the putative dusty torus, the nuclear radiation
energizes the narrow emission-line region (NLR).
Dust or electrons located within the NLR, or just outside this region, 
scatter some of the nuclear flux into our line of sight resulting
in the detection of a blue continuum and broad emission lines in
the spectrum of polarized light.
Spectropolarimetry of these AGN implies that,
from the vantage point of the scattering material, the Type~1
analogs to the narrow-line objects would be observed, effectively
unifying apparently disparate classes of objects.

With this general picture in mind, several near-IR--selected 
QSOs classified as 
Type~2 objects were also observed to search for hidden BLRs and
thereby determine if the narrow emission-line QSOs found by 2MASS
are higher-luminosity analogs of Seyfert~2 nuclei.
\citet{schmidt02} detail the results for one Type~2 2MASS QSO
that reveals a hidden BLR in polarized light and shows that for at
least some objects, the model that unifies Seyfert nuclei can be
extended to near-IR--selected QSOs. 
The sample of 2MASS QSOs selected for the current
study span nearly the full range
of AGN spectral type, as well as a large range in broadband optical
polarization.
After describing the observations (\S2) and the data obtained for individual
objects (\S3), the general trends for this optically diverse sample
of QSOs are
discussed in \S4.
We summarize our conclusions in \S5.
Primary among these are the findings
that the polarization properties of 2MASS QSOs are consistent 
with these objects being obscured by dust to various 
degrees, and that both the obscuration and the scattering that
produces the polarized flux occur near the nucleus.

\centerline{\ \ \ \ \ \ \ \ \ \ \ \ \ \ }

\section{Observations}

Optical spectropolarimetry was obtained for 21 AGN discovered by 
2MASS (Table~1) using either the 6.5~m MMT located on Mt. Hopkins,
AZ, or the 2.3~m Bok Reflector on Kitt Peak, AZ.
The data were acquired between 1999 October and 2002 July, providing
multi-epoch sampling of several objects.
All observations made use of the CCD Spectropolarimeter
\citep{schmidt92b}, upgraded with a $1200 \times 800$-pixel, thinned,
antireflection-coated, UV-sensitized CCD, and an
improved camera lens and
half waveplate.

Data were acquired with the MMT both prior to
(2001 March/April) and following the 
deposition of a high-quality aluminum coating on the primary
mirror.
The original mirror coating suffered from copper contamination
that compromised reflectivity ($R \sim 40$\% at $\lambda \sim 4000$~\AA\
and $\sim 60$\% at $\lambda \sim 8000$~\AA ) and degraded
with time.
The coating also resulted in an instrumental polarization of $\sim$1\%
that required careful calibration and removal from the 2001 March/April
data through observations of several interstellar polarization and 
unpolarized standard stars \citep{schmidt92a}.
The instrumental polarization component was reduced to $<$0.1\%
and the reflectivity increased to nearly ideal levels with the 
first fully successful primary mirror aluminization in 2001 November.

All observations utilized a 600~l~mm$^{-1}$ grating for high throughput
and wide spectral coverage, typically 4400--8800~\AA .
The full-width at half-maximum (FWHM) resolution with this grating
and a slit width of 1\farcs 1--3\farcs 0 is $\sim$17~\AA\ (3 pixels),
$\sim$800--1100~\kms , depending on telescope.
A polarimetric measurement sequence involves four separate exposures that
sample 16 orientations of the semi-achromatic half waveplate, and totals
$\sim$2000--3000~s of integration.
Typically, 2--4 such sequences were acquired for each object in a night.
Calibration of the degree of polarization ($P\/$) was made with reference to
observations of an incandescent light source through a fully polarizing
prism.
Values quoted for $P\/$ have not been corrected for
Ricean statistical bias \citep{wardle74}.
In nearly all cases this correction is inconsequential for the
data presented.
Polarization position angles were referenced to the
equatorial system by means of observations of interstellar
polarization standard stars with the identical instrumental setup. 
The spectral flux distribution is also obtained in the course
of the observations and is calibrated relative to
spectrophotometric standard stars selected from the IRAF
database \citep{massey88}.
Finally, the atmospheric O$_2$ A and B-band absorption features, as well
as the H$_2$O feature at $\sim$7200~\AA , have been removed from
the flux spectra of the AGN by observing early-type stars at nearly
the same airmass, or by scaling the results of the stellar
observations to the features seen in the AGN spectra.

Table~1 summarizes the observations by listing the observation date,
telescope, slit width, total exposure time, and the flux-weighted
mean linear polarization within the 5000--8000~\AA\ band (observed
frame; unless otherwise noted, wavelengths are given in the observer's
frame). 
In some cases more than one slit was used for an object during an
observing run as dictated by conditions.
Nine objects were observed during multiple epochs, and the broadband
mean polarization is listed for each observing run.
There is no evidence for variability of any of these objects 
between epochs or between nights of a given run.
The data for each object were averaged and these values
are also listed in Table~1.
The co-added spectropolarimetry is 
displayed in Figures~1, 2, and 3.
The coaddition of the observations was 
statistically weighted by the
data quality at each epoch.
Because of the faintness of the targets, $16 \leq B \leq 22$, the
polarization spectra can still be noisy, particularly near the
limits of the spectral coverage.

\subsection{The Sample of Objects}

The 21 objects chosen for spectropolarimetric observation
were selected from the sample 89 AGN observed by \citet{smith02}.
Seventy of these objects meet the criterion of $J - K_s > 2$
to be included in the formal sample of 2MASS red AGN \citep{cutri01}, 
and so do all but one of the spectropolarimetric targets.
Because all of the 2MASS red AGN have $M_{K_s}\/$
that fall comfortably within
the $K_s\/$-band luminosity range of optically-selected QSOs, 
\citet{smith02} classified these objects as QSOs in their
own right even though they are generally underluminous in the 
optical.
All 10 of the QSOs found to have optical broadband polarizations $>$3\%
by Smith \etal were selected for follow up spectropolarimetry.
Of these objects, spectropolarimetry of 2MASSI J151653.2+190048
(2M151653; hereafter in the text, the identification of objects
will take the form: 2M{\it hhmmss\/}, where the decimal seconds
of the J2000 Right Ascension have been truncated and the Declination has been
omitted) and of 2M165939
using the Bok Reflector have 
been reported by \citet{smith00}.
We include these objects for completeness, and in the case of
2M165939, add new data obtained with the MMT.

In addition to the highly polarized QSOs in the 2MASS sample, 
spectropolarimetry of 2M135852 is presented.
This object does not quite meet the 2MASS red AGN selection
criteria with $J - K_s = 1.8$,
but was found by \citet{smith02} to have an $R\/$-band polarization
of nearly 5\%. 

Four 2MASS QSOs
having spectra dominated by narrow emission lines
(2M100121, 2M105144, 2M130005, and 2M222554)
were observed ostensibly to search for broad emission lines in
their polarized spectra.
All of these objects have measured broadband polarizations of
$<$2\%.
The positive result for 2M130005 is reported and discussed by
\citet{schmidt02}, and these data are included here for completeness
and for comparison to other objects in the sample.
Supplementing the observations of highly polarized
and low-polarization narrow-line 
targets are six 
other 2MASS QSOs with moderate broadband polarization ($P \sim 1$--3\%).
These objects were observed as time and conditions allowed.

Except for 2M004118 (\S3.2.1), Galactic interstellar polarization (ISP)
in the sight lines
to the targets has been ignored.
The high Galactic latitude of the sample and the high
polarization of many of the objects virtually ensure that Galactic 
ISP is not a significant contributor to the observed 
polarization.
\citet{smith02} directly show this to be true for six spectropolarimetry
targets since only upper
limits can be be set for the polarization of objects near 
these sight lines with 
measurement uncertainties of $\lesssim$0.5\%.

The high quality flux spectra that are a byproduct of the spectropolarimetry
allow a re-examination of the optical spectral classification of the 
2MASS AGN that was based on the original,
confirming spectroscopy \citep{cutri01}
listed in \citet{smith02}.
We use the AGN classification guidelines of \citet{ho97} that were
also used by Cutri \etal in generating the 2MASS AGN sample.
This scheme is very close to the classification criteria of
\citet{veilleux87}.
Some of the original classifications were ambiguous because H$\beta\/$
was not available for various reasons.
This emission line is clearly identified in all of the 21 
spectropolarimetry targets and allows for a more definitive classification 
of four objects.
2M135852 and 2M150113 are now classified as Type~1 AGN and 2M163700
(see Figure~2) is Type~1.5 owing to the prominent narrow-line
component of H$\beta\/$.
\citet{schmidt02} reclassified 2M130005 as a Type~2 AGN.

Four other objects were reclassified based on the new spectra.
2M105144 is Type~1.9 as opposed to the original LINER classification
because it possesses a broad  H$\alpha\/$ component (Figure~3).
2M222221 (Figure~2)
is reclassified as Type~1.5 because of the strong, narrow emission-line
component seen in the Balmer-line profiles and the relatively
high ratio of [\ion{O}{3}]$\lambda$5007 to H$\beta\/$ flux.
In contrast, the smaller [\ion{O}{3}]$\lambda$5007/H$\beta\/$
flux ratio, absence of a distinguishable narrow emission-line
component for H$\beta\/$, and very strong optical \ion{Fe}{2} features 
favor classifying 2M230307 as Type~1.
The original Starburst classification is somewhat problematic for 
2M100121. 
The emission-line ratios are consistent with either Starburst or
Seyfert~2 designation, but given its high QSO-like luminosity
in the near infrared, it is difficult to imagine
that hot stars power the nuclear emission.
Therefore, we reclassify 2M100121 as a Type~2 AGN.
Table~1 summarizes the adopted AGN spectral classifications for
the spectropolarimetry sample.

\section{Results}

\subsection{Spectropolarimetric Measurements}

The spectropolarimetric results are presented in Figures~1--3.
For all objects, the sequence of four panels depicts, from top to bottom:
the polarization position angle $\theta\/$, the rotated Stokes
parameter $q'\/$ for a coordinate system aligned with the 
mean polarization of the source in the 5000--8000~\AA\
bandpass ($q' = q \cos 2\theta + u \sin 2\theta\/$), the polarized
(or ``Stokes'') flux $q' \times F_\lambda\/$, and finally, the
total spectral flux $F_\lambda\/$.
Two artifacts should be noted.
First, even though the data have been corrected for terrestrial
absorption features, the O$_2$ and H$_2$O bands are sufficiently
deep that increased noise can be noted in the polarized flux
spectra around the wavelengths of these features.
Second, fringing 
sharply modulates the CCD quantum efficiency for $\lambda \gtrsim 8000$~\AA .
When combined with a small amount of flexure in the instrument, the 
result is that measured quantities show rapid oscillations
as a function of wavelength around their mean values for a few objects.
2M010607 and 2M135852 (Figure~1) show examples of this effect.

Objects have been divided into three groups: Type~1, Type~1.5,
and Types~1.8--2 to compare and contrast characteristics over
a wide range of emission-line properties.
Summarized in Table~2 are the equivalent width (EW in \AA )
and line flux (in units of 10$^{-14}$~erg~cm$^{-2}$~s$^{-1}$)
for H$\beta\/$,
H$\alpha\/$ (for objects with $z < 0.34$), and [\ion{O}{3}]$\lambda$5007.
The line widths of H$\alpha\/$ and H$\beta\/$ (FWHM in
\kms ) are also given.
Because the emission lines from the NLR are
generally unresolved, we do not list the [\ion{O}{3}]$\lambda$5007
line width, but instead tabulate the [\ion{O}{3}]$\lambda$5007
luminosity ($H_0 = 75$~\kms~Mpc$^{-1}$, $q_0 = 0$, and $\Lambda = 0$
are assumed throughout).
Measurements for all of the available quantities are given for both
the total flux spectrum ($F_\lambda\/$) and the polarized flux
spectrum ($q' \times F_\lambda\/$) in the observed frame.
A colon after an entry signifies that the measurement is difficult and the
reason for the resulting uncertainty
is identified in the last column
of Table~2.

Measurements of H$\alpha\/$ are contaminated to various degrees by
[\ion{N}{2}]$\lambda\lambda$6548,6583.
The contribution to the H$\alpha\/$ flux by [\ion{N}{2}] is not
large except for the Type~1.8--2 QSOs, where the [\ion{N}{2}]
lines may emit $\gtrsim$1/2 of the flux and broaden the feature.
Flux from [\ion{S}{2}]$\lambda\lambda$6717,6731 was excluded
from measurements of H$\alpha\/$ + [\ion{N}{2}].
Given the spectral resolution of the observations, the values listed in 
Table~2 for H$\alpha\/$ include the contributions from
both the narrow and broad emission-line components as well as
[\ion{N}{2}].
Likewise, the measurements of H$\beta\/$ include both broad- and narrow-line
components of the emission feature.

Optical \ion{Fe}{2} emission is detected in all of the Type~1 and
1.5 objects except 2M165939.
Since measurements of H$\beta\/$ and
[\ion{O}{3}]$\lambda$5007 are affected by strong
multiplets in this wavelength region,
the \ion{Fe}{2} emission was removed from
$F_\lambda\/$ and $q' \times F_\lambda\/$ (if detected in polarized 
light) before these lines were measured.
An optical \ion{Fe}{2} template based on the narrow-line QSO I~Zw~1
\citep{boroson92}
was broadened to the approximate H$\beta\/$ FWHM, scaled to
the estimated flux of the feature observed at $\sim$4450--4750~\AA\ 
(rest frame),
and subtracted from the 2MASS QSO spectra.

The continuum properties of the 2MASS QSOs are characterized
in Table~3.
For more than half of the sample, a significant contribution
to the observed flux is made by starlight originating in the host
galaxy of the QSO.
We have estimated the amount of starlight falling within the observing 
aperture for all objects showing stellar absorption features in
their spectra.
For three other objects that do not show obvious stellar features, an
estimate of the host galaxy flux is provided by high resolution
imaging obtained by \citet{marble03}
using the Wide Field/Planetary Camera~2 (WFPC2) and F814W filter aboard
the {\it Hubble Space Telescope\/} ({\sl HST\/}).
The elliptical galaxy spectrum of NGC~3379 \citep{kennicutt92} was used
as a template to extrapolate the measured stellar contribution within
the F814W filter to shorter wavelengths.
This same template also reasonably accounts for the observed
absorption features of the 11 objects where the host galaxy
is directly detected in the total flux spectra.
The ratio of starlight from the host galaxy to total flux
at 5500~\AA\ in the rest frame is listed in Table~3, along
with the polarization of the remaining light in the
observed 6000--7000~\AA\ band after the
subtraction of the assumed unpolarized stellar component.

Power-law fits to the continua for both the polarized flux spectrum
and the host galaxy-subtracted 
total flux spectrum were made to characterize the continuum properties
of the sample.
The power-law index, $\beta\/$, where $F_\lambda \propto \lambda^\beta$,
is given in Table~3.
The power-law fits avoid regions of exceptionally high noise,
major emission lines, and obvious \ion{Fe}{2} features if present.

Following $\beta\/$ in Table~3 are the strengths of [\ion{O}{3}]$\lambda$5007,
H$\alpha\/$, and \ion{Fe}{2} relative to the H$\beta\/$ flux. 
If possible, entries for these line ratios are listed for both the total
and polarized flux spectra.
As in Table~2, the measurement of $F_{{\rm H}\alpha\/}$ includes flux
from [\ion{N}{2}]$\lambda\lambda$6548,6583.
The strength of the optical \ion{Fe}{2} emission is characterized by the
its flux within the 
rest frame 4450--4750~\AA\ band.

Finally, the last three columns in Table~3 list the line and continuum
polarizations for H$\beta\/$ and H$\alpha\/$ and an estimate of the 
polarization of the NLR based on the measurements of
[\ion{O}{3}]$\lambda$5007.
The polarizations of the permitted lines are calculated using the
measured total and polarized line fluxes.
The continuum polarization at H$\alpha\/$ and H$\beta\/$ is derived from the
power-law fits to the continua described above.
Throughout this paper, we assume that the generally low observed
polarization of the NLR is well-represented
by the strength of narrow-line features in the polarized flux spectrum.

\subsection{Type 1 Objects}

About 3/4 of the 2MASS red AGN sample show broad emission lines
in their total flux spectra \citep{cutri01}.
In the spectropolarimetric sample,
Type~1 objects 
correspond to those with
$F_{\rm [O~III]}$/$F_{{\rm H}\beta} < 0.5$.
Spectropolarimetry of seven Type~1 2MASS QSOs and one highly polarized,
broad emission-line
AGN found by 2MASS (2M135852) is shown in Figure~1.

\subsubsection{2MASSI J004118.7+281640}

Optical unfiltered polarimetry of 2M004118 yields
$P = 2.2\% \pm 0.3$\% at $\theta$ = 104\dgr$\pm$4\dgr\  
\citep{smith02}.  
Our spectropolarimetry confirms the level of polarization, and
we measure a flux-weighted polarization position angle of 97\dgr\ 
in the 5000--8000~\AA\ bandpass.
A portion of this originates as Galactic ISP, since 
\citet{smith02} found 
$P = 0.74$\%, $\theta = 117$\dgr\ for a nearby field star.
The data displayed for 2M004118 in Figure~1 and Tables 2 and 3
have been corrected for this ISP,
assuming that it follows the standard Serkowski law
\citep{serkowski75}
with a maximum polarization of 0.74\% at 5500~\AA .

Although the polarization of 2M004118 is diminished by the ISP
correction, it is clear that the
object is intrinsically polarized. 
The spectrum of $q'$ shows that the H$\alpha$, H$\beta$,
and [\ion{O}{3}] emission lines are not polarized to the same degree
as the surrounding continuum, as would be expected if ISP were
the sole polarizing mechanism.
The ISP correction yields a mean polarization in the
5000--8000~\AA\ bandpass
for 2M004118 of $P \sim 1.7$\% at $\theta \sim 89$\dgr .

2M004118 is one of the bluest objects in the sample, with
$B - K_s = 3.4$ and $\beta_{F_\lambda} = -2.2$, and has strong
optical \ion{Fe}{2} emission features.
The polarized flux spectrum shows a continuum nearly as blue
as the total flux spectrum.
Hydrogen~$\alpha$ exhibits changes in $q'\/$ and $\theta\/$
across its profile.
The core of the line marks a transition between the red wing
of the line that is polarized to the degree and position angle
of the continuum,
and the less polarized blue wing.
The position angle swings through $\sim$90\dgr\ 
near the line core.
A similar signature in $q'\/$ is identified for H$\beta$
and the strong \ion{Fe}{2} feature blended with [\ion{O}{3}]$\lambda$5007
at 5980~\AA .
The complex structure of the H$\alpha$ polarized flux profile is
reminiscent of some highly polarized Seyfert 1 nuclei
\citep[e.g.,][]{goodrich94,smith95,smith97,martel97,martel98,smithj02}.

\subsubsection{2MASSI J010607.7+260334}

In contrast to 2M004118, 2M010607 is a much redder object optically
($\beta_{F_\lambda} = 1.3$; $B - K_s > 6$), more than 3~mag fainter,
and more highly polarized.
2M010607 shows \ion{Fe}{2} emission features nearly as strong relative
to H$\beta$ as in 2M004118.
Unfortunately, the redshift of 2M010607 places H$\alpha$ outside
the sensitivity range of the spectropolarimeter.
The primary feature of the polarized flux spectrum is the presence
of broad ($\sim$1900~\kms\ FWHM) H$\beta$ with roughly the same 
equivalent width as seen in the total flux spectrum.
There is no evidence of [\ion{O}{3}]$\lambda$5007 in the polarized flux
spectrum and the polarized continuum is quite red
($\beta_{q' \times F_\lambda} = 0.5$).

\subsubsection{2MASSI J125807.4+232921}

Spectropolarimetry of 2M125807 confirms the low $R\/$-band polarization
measurement reported in \citet{smith02}.
The object is $\sim$1\% polarized and the appropriate quantities listed in
Table 3 have been measured after subtracting an elliptical
host galaxy spectrum that accounts for 13\% of the total light received
at 5500~\AA\ in the rest frame.
Because stellar spectral features are not apparent in the spectrum,
the contribution of the host galaxy was estimated from an {\sl HST\/}
WFPC2 F814W image \citep{marble03}.
In this case, the host galaxy has little impact on the measured
polarization and other quantities.

The spectral slopes of total flux and polarized flux continua
are similar to those of 2M004118, as is the Balmer decrement
in the total flux spectrum.
Subtraction of the \ion{Fe}{2} template from the total flux spectrum
accounts for all of emission peaks between H$\beta\/$ and $\sim$7000~\AA .
There is no evidence of [\ion{O}{3}] emission from 2M125807, making this
the only object in spectropolarimetric sample not exhibiting emission lines
from this species.
No definitive features are seen in the polarized flux spectrum.

\subsubsection{2MASSI J132917.5+121340}

Although 2M132917 is polarized only $\sim$1\% around H$\alpha$, its
polarization rises rapidly to the blue.
Indeed, the object exhibits one of the bluest polarized flux spectra
of this sample ($\beta_{q' \times F_\lambda} = -2.0$),
much bluer than the total flux
continuum ($\beta_{F_\lambda} = -$0.5).
The feature seen in the polarized flux spectrum near [\ion{O}{3}]$\lambda$5007
is much narrower ($\sim$1~pixel) than the instrumental resolution
and could simply be a noise spike due to a cosmic ray.
The flux in this narrow feature is used as an upper limit to the
polarized flux from
[\ion{O}{3}]$\lambda$5007.
Comparison of the H$\beta$/[\ion{O}{3}]$\lambda$5007 flux ratio for
both $F_\lambda\/$ and $q' \times F_\lambda\/$ suggests that the polarization
of the BLR is at least four times higher than that of the NLR.

The Balmer lines reveal a pronounced red asymmetry in their profiles in
the total
flux spectrum.
Much like 2M004118, the H$\alpha$ polarized flux is emitted in the
red wing of the line. 
Near the line core and in the blue wing $q'\/$ falls well below 0.
Since there is no corresponding feature seen in the rotated
$u\/$ Stokes parameter, the polarization of the blue wing of
H$\alpha$ is orthogonal to that of the red wing.
The two polarized components would then tend to cancel near the line core,
as seen in the
$q'\/$ and $q' \times F_\lambda\/$ panels for 2M132917 in Figure~1.
The line center of H$\beta$ in polarized flux is
shifted $\sim$1000~\kms\ to the red
of the line peak in total flux.
This could be caused by the 
same polarization structure seen in H$\alpha$,
although the spectropolarimetry does not have the
signal-to-noise ratio
(S/N) to clearly
show these features.

\subsubsection{2MASSI J135852.5+295413}

This object is not a member of the 2MASS red QSO sample because
$J - K_s = 1.8$ in revised 2MASS photometry.
Nevertheless,
2M135852 has a $K_s\/$ luminosity well within the range of the 
2MASS QSOs and $B - K_s = 4.15$, redder than the median $B - K_s\/$
color index of the Type~1 objects observed by \citet{smith02}.
In addition, $R$-band imaging polarimetry shows $P \sim 4.8$\%,
resulting in the object being added to the list of 2MASS AGN for
follow up observation.

Spectropolarimetry confirms the highly polarized nature of 2M135852.
The polarization reaches $\sim$8\% at 4600~\AA\ and has a position
angle that is constant with wavelength.
The spectrum reveals stellar absorption features
that can be removed with a starlight fraction of
$\sim$60\%
in the observing aperture at 5500~\AA\ (rest 
frame).
This implies an intrinsic
polarization in the 6000--7000~\AA\ band double of that observed,
and well over 10\% for $\lambda < 5000$~\AA .
The host galaxy also leads to higher observed polarization in
the Balmer lines than in the continuum since the
relative dilution in the lines is reduced.
Balmer line widths in $F_\lambda$ and $q' \times F_\lambda$ are
about the same. Even with the correction to $F_\lambda$ for 
a red stellar component, the polarized flux continuum is bluer
than the nuclear light in the total flux spectrum.

The $q'\/$ panel in Figure~1 for 2M135852 indicates that the 
polarization at [\ion{O}{3}]$\lambda$5007 is much diminished from the
surrounding continuum.
Estimating the [\ion{O}{3}]$\lambda$5007 flux in the polarized spectrum
yields a polarization for the NLR of $\lesssim$2.5\%.
Unlike several other Type~1 objects, 2M135852 shows no structure
in $\theta\/$ across H$\alpha$ in the polarized flux spectrum.

\subsubsection{2MASSI J150113.1+232908}

2M150113 is optically the reddest of the eight Type~1 objects with
$\beta_{F_\lambda} = 2.3$,
and the \ion{Ca}{2}~H and K break is seen in its spectrum.
The stellar features imply that the host galaxy contributes $\sim$60\%
of the light at 5500~\AA\ (rest frame) within the
$1\farcs 5 \times 3\farcs 8$
observation aperture.
This, in turn, suggests that the polarization of the nuclear light
in the $R\/$-band is $\sim$8\%, in contrast to the $P = 3$--4\% measured
by the spectropolarimetry and by \citet{smith02}.
The increased polarization seen at
H$\alpha\/$ and H$\beta\/$
is explained by unpolarized starlight included in the aperture,
as is also the case for 2M135852.

The noise level of the polarized flux spectrum makes
it is difficult to determine if [\ion{O}{3}]$\lambda$5007 is present in
polarized light. 
An upper limit to the [\ion{O}{3}]$\lambda$5007 polarized flux suggests
that the emission from the
NLR cannot be polarized more than $\sim$1/3 of the polarization
measured for the continuum and BLR.

The degree of polarization of 2M150113
rises to the blue as indicated by the spectral
index of the polarized continuum relative to the total
nuclear continuum (after the 
host galaxy spectrum has been subtracted).
In addition, the Balmer decrement is large in both the total flux
spectrum ($F_{{\rm H}\alpha}/F_{{\rm H}\beta} \sim 12$) and the
the polarized flux spectrum ($F_{{\rm H}\alpha}/F_{{\rm H}\beta} \sim 7$).
The large decrement in $q' \times F_\lambda\/$ indicates that
that even the light scattered into our view is substantially reddened.

\subsubsection{2MASSI J151653.2+190048}

\citet{smith00} describe the polarization properties of this object.
We include it here to compare with the rest of the sample and
to tabulate the measurements listed in Tables~2 and 3.
The object is the most luminous at $K_s\/$ of the Type~1 QSOs in the
spectropolarimetry sample.

The strong optical \ion{Fe}{2} emission features are observed in the
spectrum of polarized flux at about the same strength
relative to H$\beta$ as in the 
total flux spectrum.
The only other QSO in this sample to unambiguously show polarized
\ion{Fe}{2} features is 2M091848 (see \S3.3.1).
The broad emission lines are polarized at about the same level
as the continuum, which exhibits a strong rise in polarization
from 8600~\AA\ ($P \sim 7$\%) to 4600~\AA\ ($P \sim 14$\%).
The bluer polarized flux spectrum relative to the total flux
spectrum is accompanied by a Balmer decrement in polarized light
nearly a factor of 2 smaller than for the total flux spectrum.

The NLR is unpolarized as indicated by the absence of
[\ion{O}{3}]$\lambda\lambda$4959,5007 in the polarized flux
spectrum.
Structure is
observed in $q'\/$ and $\theta$ across H$\alpha\/$, and to
a lesser degree, H$\beta\/$ (Figure~1).  
The red wings of the Balmer lines are more highly polarized than the
blue wings.
This same signature is seen for H$\alpha\/$ in the Type~1 objects 2M004118
and 2M132917.

\subsubsection{2MASSI J230307.2+254503}

This object is the strongest \ion{Fe}{2} emitter relative to H$\beta\/$ of the 
QSOs in this sample.
Unfortunately, H$\alpha\/$ falls very close to the red end of the
spectrum, making measurements of this emission line somewhat problematic.
The S/N of the spectropolarimetry is too low around the H$\alpha\/$
line for any meaningful measurements.
Despite the faintness of this QSO and its relatively low optical
polarization, H$\beta\/$ can be seen in emission in polarized
flux.
Another feature apparent in Figure~1 is the decline of polarization to
$q' \sim 0$ at [\ion{O}{3}]$\lambda$5007.

\subsection{Intermediate Objects}
 
In this section we discuss QSOs classified as Type~1.5.
The narrow line-dominated
Type~1.8 and 1.9 objects, also considered ``intermediate'' AGN
spectral types,
have been lumped together with the Type~2 2MASS QSOs (\S 3.4).
This division also separates objects in the sample by their
[\ion{O}{3}]$\lambda$5007/H$\beta\/$ flux ratios, with
Type~1.5 QSOs in this sample having
$0.5 \leq F_{\rm [O~III]}/F_{{\rm H}\beta} \leq 3$.
Figure~2 displays the results for these objects.

\subsubsection{2MASSI J091848.6+211717}

The polarization of 2M091848 increases strongly to the blue, reaching
over 10\% for $\lambda < 5000$~\AA , and has a position angle that is
constant with wavelength.
The Balmer lines, including H$\gamma\/$, are seen in polarized flux, as
is [\ion{O}{3}]$\lambda$5007.

2M091848 along with 2M151653 are the only two objects to clearly
show polarized optical \ion{Fe}{2} emission features.
The \ion{Fe}{2} emission on either side of H$\beta\/$ in 
2M091848 is more prominent
in polarized flux than in the total flux spectrum
largely because of the significant amount of host galaxy starlight
included in the total flux.
Stellar absorption features in $F_\lambda\/$ suggest
a stellar--to--total light ratio at 5500~\AA\ (rest frame) of $\sim$0.4.
This contribution by the host galaxy implies
that the intrinsic $R\/$-band polarization of the QSO is $\sim 10$\%,
as opposed to $6.3 \pm 0.1$\% measured by \citet{smith02} without
correction for starlight.

The polarized flux spectrum of 2M091848 shows [\ion{O}{3}]$\lambda$5007,
though its strength
relative to H$\beta\/$ is much reduced
and the weaker [\ion{O}{3}]$\lambda$4959 line does not rise above the
S/N level in polarized light.
The [\ion{O}{3}]$\lambda$5007 measurements yield a polarization for this
line of $\sim$2\%, which is a factor of 4--6 lower than the 
polarization of H$\beta\/$ and the continuum at this wavelength.

\subsubsection{2MASSI J134915.2+220032}

2M134915 is the least luminous QSO in the sample
in the near infrared ($M_{K_s} = -24.9$) and
has the smallest near-IR-to-optical flux ratio inferred 
from $B - K_s\/$.
Although the observed polarization for 2M134915 is much lower than 
in 2M091848, many of the polarization properties of these two QSOs
are similar.
Both exhibit redder continua in total
flux than in polarized flux.
Both objects clearly show a polarized [\ion{O}{3}]$\lambda$5007 line.
In 2M134915, [\ion{O}{3}]$\lambda$4959 is also seen in $q' \times F_\lambda\/$.
Based on the [\ion{O}{3}] measurements, the NLR is polarized by $<$1\%;
$<$1/4 the polarization of the continuum at this wavelength.

Unlike most of the Type~1 objects, which typically show Balmer lines
in polarized light that have about same FWHM as in total flux,
polarized H$\beta\/$ in 2M134915
has a width $\sim$2700~\kms\ larger than measured in the total flux
spectrum. 
To a large extent, this reflects the lessened contribution from
the NLR in the polarized H$\beta\/$ line profile compared to that for the
line in total flux.
The increased polarization seen in the Balmer lines is caused by the
substantial contribution from the host galaxy within the observing aperture.
From the stellar features observed in the spectrum, nearly
half of the light within the aperture at 5500~\AA\ (rest frame)
is starlight from
the host galaxy.

\subsubsection{2MASSI J163700.2+222114}

The \ion{Ca}{2}~H and K break is readily apparent in the spectrum 2M163700.
The stellar features imply a flux contribution by the host galaxy of
$\sim$0.5 at 5500~\AA\ (rest frame) that in turn suggests an intrinsic
level of polarization of the AGN of over 4\%.
Although the continuum slope of the polarized flux spectrum is bluer
than that of the total flux, both are red
($\beta_{F_\lambda} \sim 2.0$; $\beta_{q' \times F_\lambda} \sim 0.9$).
Large Balmer decrements in both polarized and total flux spectra
further confirm the highly reddened nature of this QSO.
In fact, these decrements are by far the largest for the Type~1.5
QSOs, and only 2M150113 challenges 2M163700 in this parameter among
the Type~1 objects.
There is no evidence for [\ion{O}{3}] (or [\ion{O}{2}]$\lambda$3727) in the 
polarized flux spectrum.

\subsubsection{2MASSI J165939.7+183436}

Spectropolarimetry of this object obtained at the Bok Reflector 
is reported and discussed by \citet{smith00}.
Subsequently,
an MMT observation of similar quality was obtained of 2M165939 on 2001 March 31.
No significant differences are seen between these observations and
the co-added results from the two telescopes are displayed in Figure~2.
Like 2M134915 and 2M163700, the host galaxy of 2M165939 contributes
to the observed flux.
Various slit widths (1\arcsec\/--3\arcsec ) were employed for the observations
of 2M165939
and a starlight-to-total flux ratio of 0.26
in the rest frame $V\/$-band is adopted for the composite spectrum.
This ratio
is based on the identification of a weak absorption feature at
$\sim$6060~\AA\ with \ion{Mg}{1}~b.

Narrow emission lines seen in the total flux spectrum
are not seen in polarized light. 
In contrast, H$\alpha\/$ and H$\beta\/$ are prominent
in the polarized spectrum and are much broader than in either
$F_\lambda\/$ or $q' \times F_\lambda\/$ for any of the Type~1 objects.
The asymmetric profile of the broad component of H$\alpha\/$ in
the total flux spectrum is somewhat reproduced in polarized light with
the blue wing of the line being sharper than the red wing.
The broad hump identified with H$\beta\/$ in the polarized spectrum
has an impressive FWHM of nearly 16,000~\kms .

\subsubsection{2MASSI J170003.0+211823}

2M170003 has the highest redshift, $z = 0.596$, of the 2MASS QSOs
and is one of the reddest objects
in the sample.
It also shows the highest observed optical broadband polarization
\citep[$P \sim 11$\%;][]{smith02}.
The spectropolarimetry confirms the high level of polarization and
reveals that the polarized spectrum is about as red as the extremely red
continuum observed in total light.
Again, the major difference between the spectra of total and polarized 
light is that the emission from the NLR region is absent in the
polarized flux spectrum.
Broad H$\beta\/$ and H$\gamma\/$ are detected in polarized spectrum
and are polarized to about the same degree as the continuum.
The position angle is constant across the entire 
spectrum.

\subsubsection{2MASSI J222202.2+195231}

The highly polarized QSO 2M222202 is the most luminous ($M_{K_s} = -28.6$)
of the 
sample of 70 2MASS red QSOs observed by \citet{smith02} and it displays
a very rich emission-line spectrum that includes emission lines
of H, \ion{He}{1}, \ion{He}{2}, \ion{O}{3},
[\ion{O}{2}], [\ion{O}{3}], [\ion{Ne}{3}],
[\ion{Ne}{5}], and [\ion{S}{2}] (Figures~2 and 4).
Weak optical \ion{Fe}{2} features are also present.
Remarkably, with the possible exception of \HB ,
none of these features are seen in the polarized flux spectrum.
The rotated Stokes parameter shows complex
structure across the spectrum
with dramatic decreases in polarization at each emission line.
Indeed, significant decreases in $q'\/$ are even seen at the locations of
lines with small equivalent widths, such as \ion{He}{2}~$\lambda$4686 and
[\ion{O}{2}]$\lambda$3727.
The continuum polarization increases from $\sim$10\% at the red
end of the spectrum to nearly 18\% at $\sim$5370~\AA ,
in the gap between H$\epsilon\/$
and [\ion{Ne}{3}]$\lambda$3869 + \ion{He}{1}~$\lambda$3889 +
H$\zeta\/$.
The continuum polarization then decreases from this point until the
blue end of the observed spectrum at 4200~\AA\ where $q'\/$ is 
again $\sim$10\%.

The position angle spectrum is not as dramatic
as $q'\/$, but is of sufficient S/N to show interesting structure
as well (Figure~4).
Generally from $\sim$5600~\AA\ to 8600~\AA , $\theta\/$ is nearly 
constant at around 120\dgr .
However, a 10\dgr\ rotation can be seen at the location of
[\ion{O}{3}]$\lambda$5007 and more tentatively in
the core of \HB\ and at
\HC\ + [\ion{O}{3}]$\lambda$4363.
A smaller rotation in the same sense can be discerned at
[\ion{O}{3}]$\lambda$4959.
Although $q'\/$ is based on a position angle of 120\dgr ,
these $\lesssim$10\dgr\ rotations in $\theta\/$ are too small
to significantly affect the spectrum of $q'\/$ presented in
Figure~4.

In addition to the discrete rotations in $\theta\/$ in some of the
emission lines, a broad
feature is observed centered near 5000~\AA .
The rotation has about the same amplitude as
seen for [\ion{O}{3}]$\lambda$5007, and although the S/N is
much diminished at the blue end of the spectrum, $\theta\/$
appears to recover back to $\sim$120\dgr\ by 4400~\AA .
This feature coincides with the decrease in
the continuum polarization in the blue and the emergence of
the ``3000~\AA\ bump'' in the flux spectrum
\citep[see e.g.,][]{wills85}.
At these wavelengths, the 3000~\AA\ bump is primarily emission from
high-level Balmer lines and the Balmer continuum.
In 2M222202, this
feature, [\ion{O}{3}]$\lambda\lambda$4959,5007, and other emission lines
must be polarized since $\theta\/$ cannot be affected
by unpolarized light. 
The continuum is polarized to a much higher degree than the emission from
the NLR since the rotation in $\theta\/$
at strong NLR features is only $\sim$10\dgr .
Measurement of the [\ion{O}{3}]$\lambda$5007 line implies a polarization
for the NLR light of $P \sim 0.8$\% at $\theta \sim 85$\dgr .
Assigning the same polarization to the 3000~\AA\ bump
does not readily account for the polarization observed near the Balmer
limit, although the decomposition of the spectrum into two polarized
components is highly dependent on uncertain choices of the
continuum strength and
polarization in this spectral region.
To be consistent with the results at [\ion{O}{3}]$\lambda$5007, either
the 3000~\AA\ bump must be a larger contributor to the total flux
around 5000~\AA\ than implied by a simple extrapolation
from longer wavelengths, or the continuum polarization at $\sim$5000~\AA\
is not as high as indicated by the rapid rise in $P\/$ observed at
longer wavelengths.

The deviations in $\theta\/$ across the spectrum make it more 
difficult to interpret the spectrum of polarized light.
The emission lines seen in $F_\lambda\/$ are absent in $q' \times F_\lambda\/$.
There is an abrupt break in the polarized flux spectrum around
H$\beta\/$/[\ion{O}{3}]$\lambda\lambda$4959,5007 where the continuum
flattens from $\beta_{q' \times F_\lambda\/} \sim -2.3$ for 
$\lambda \gtrsim 6800$~\AA\ to $\beta_{q' \times F_\lambda\/} \sim -0.4$ for
$\lambda \lesssim 6800$~\AA .
A very low contrast emission feature is tentatively identified at
the location of H$\beta\/$ and [\ion{O}{3}]$\lambda\lambda$4959,5007.
If this feature is actually polarized flux from H$\beta\/$, the line is
polarized at only $\sim$1/2 the level of the continuum.
An identification with H$\beta\/$ also implies that the scattered
line profile is even broader ($\sim 17,000$~\kms\ FWHM) than the polarized
H$\beta\/$ profile observed in 2M165939 (\S 3.3.4).
The ``feature'' could also simply be a manifestation of the position angle
rotation seen in the emission lines.
We have included measurements of the feature in Tables~2 and 3 under
the assumption that it is H$\beta\/$.

The ambiguity in the identification of polarized H$\beta\/$, and the lack of
other lines in the polarized flux spectrum of 2M222202, results in 
greater uncertainty in identifying the mechanism responsible for the 
high continuum polarization.
The polarization of the narrow emission-lines must either be caused by
scattering from material far enough away from the NLR to 
result in a net polarization of the light, or dichroic absorption 
in the sight line between us and the NLR of 2M222202.
Dichroic absorption is essentially ruled out as the cause of the
continuum polarization because an enormous amount of extinction would
be required to account for polarization approaching 20\%.
Scattering by dust or electrons requires that the particles
be located very close to the nuclear continuum source.  
To account for the very weak (or nonexistent) broad-line features in
the polarized flux, the scatterers need to be intermixed
with the gas in the BLR, or located just exterior to the line-emitting
region.
Polarized flux produced in such close proximity to the BLR
probably favors electrons
as the scatterers since the environment is likely to be too harsh for the
survival of dust grains (however, see e.g., \citealt{goodrich95}
for evidence that dust can exist in the BLR).

If indeed the polarized flux spectrum of 2M222202 is featureless, 
synchrotron radiation is a possible source of the highly polarized continuum.
In this case, the optical flux and
polarization would be expected to strongly vary  
as observed for OVV quasars
and BL~Lacertae objects.
In fact, there is no evidence for variability in 2M222202 
from the three epochs of spectropolarimetry obtained between 1999
September and 2002 July.
2M222202 is detected as a 5~mJy radio source at 1.4~GHz in the
NRAO VLA Sky Survey \citep{condon98},
but the object is
$\sim 100-1000\times$ less luminous in the radio
than objects that show strong and variable optical synchrotron continua.

\subsubsection{2MASSI J222221.1+195947}

The polarization of 2M222221 is only $\sim$1\% even after subtracting the
host galaxy contribution to the observed spectrum estimated from {\sl HST\/}
imaging
\citep[][$F_{\rm gal}/F_{\rm Total} \sim 0.3$ in
the rest frame $V\/$-band]{marble03}.
Despite the low level of polarization, broad H$\alpha\/$ and H$\beta\/$
are apparent in $q' \times F_\lambda$.
The lines in polarized flux are
blue-shifted by $\sim$1000~\kms\ relative
to the narrow Balmer-line components in the total flux spectrum.
Also detected in the polarized flux is [\ion{O}{3}]$\lambda$5007, although
at a reduced strength relative to H$\beta\/$.
This yields
an estimate of the NLR polarization of only $\sim$0.3\%.
One other feature of note for 2M222221 is a large rotation
in $\theta\/$ across H$\alpha\/$.
The position angle is $\sim$160\dgr\ in the blue wing of the line, 
$\sim$180\dgr\ around the line core, and then rotates to 
130\dgr --110\dgr\ in the red wing.

\subsection{Type 1.8, 1.9, and 2 Objects}

The objects described in this section would traditionally not
be classified as QSOs on the basis of their optical
spectra since their narrow emission
lines far outshine any weak broad permitted lines.
Indeed, their H$\beta\/$ widths are $\sim$1000~\kms\ (FWHM) or less.
All but one show stellar absorption features 
with equivalent widths suggesting that well over half 
of the optical continuum flux is starlight from the galaxy hosts.
The 2MASS results tell a much different story for these red QSOs:
all have $M_{K_s} < -25.5$ and have large near-IR/optical
flux ratios ($B - K_s > 5$).
Spectropolarimetry of the Type~1.8--2 QSOs is displayed in Figure~3.

Combining the Type 1.8--2 objects separates QSOs
with $F_{\rm [O~III]}/F_{{\rm H}\beta} > 3$ from the rest of the 
sample.
Their total flux spectra generally have larger Balmer decrements
than those measured for the Type~1s and 1.5s.
This is due in part to the
inclusion of [\ion{N}{2}]$\lambda\lambda$6548,6563 in the spectral
measurements of H$\alpha\/$, since
the [\ion{N}{2}] flux in several objects is comparable to H$\alpha\/$.
The presence of the [\ion{N}{2}] lines also yields
widths in total flux that are roughly
double those measured for H$\beta\/$.
Optical \ion{Fe}{2} emission is not detected in these six objects.

\subsubsection{2MASSI J010835.1+214818}

2M010835 is the only object besides 2M171559 (\S3.4.5)
among the 15 Type~1.8--2 AGN in the 2MASS red
QSO sample that shows broadband optical polarization $>$3\%
\citep{smith02}.
Spectropolarimetry confirms that the polarization of the continuum is
$\sim$5\% and reveals stellar features in the total flux spectrum.
The strength of the features suggests that about 80\% of the 
continuum at $\sim$7000~\AA\ is starlight, requiring that the true continuum
polarization of the AGN be over 20\% redward of
\ion{O}{3}$\lambda$5007 and
$<$10\% at the blue end of the spectrum.
However, given the faintness of the continuum, this estimate is
very uncertain.
A fainter host galaxy would, of course, result in lower overall
polarization as well as a
smaller relative decrease in the polarization in the blue
since the spectrum of the AGN 
would be redder.

Permitted and forbidden lines are observed in the polarized flux
spectrum.
Although H$\alpha$ is near the red end of the observed spectrum where
the S/N is low, the line (plus possibly [\ion{N}{2}]) can be seen in polarized flux.
Both [\ion{O}{3}] lines are apparent in $q' \times F_\lambda\/$, but at
much smaller equivalent widths than in total flux.
The ratio of the polarized-to-total
[\ion{O}{3}] flux implies a polarization of $\sim$1.5\%, substantially
less than for the continuum.
Relative to [\ion{O}{3}], the Balmer lines are stronger in polarized light 
than in $F_\lambda\/$, indicated a higher polarization for the 
permitted lines.

\subsubsection{2MASSI J100121.1+215011 and 2MASSI~J222554.2+195837}

The Type~2 QSOs 2M100121 and 2M222554 are polarized at a very low level ($<$1\%)
and no features can be discerned in the spectrum of polarized light.
Stellar absorption features in the total flux spectra, including
the \ion{Ca}{2} break, indicate that the host galaxy contributes
$\sim$80\% of the light from these objects at 5500~\AA\ (rest frame) 
within the 1\farcs 1$\times\sim$4\arcsec\ apertures employed for the
spectropolarimetry.
Correcting the data for even this large amount
of unpolarized flux only elevates the polarization of 2M222554 to
$\sim$1\%.

\subsubsection{2MASSI J105144.2+353930}

2M105144, like 2M100121 and 2M222554,
is not highly polarized and exhibits prominent
stellar features in its spectrum.
The ratio of starlight to AGN light in the 1\farcs 5$\times$4\farcs 6
aperture used is estimated to be $\sim$2.3 in the rest-frame $V\/$-band.
Subtraction of the assumed elliptical galaxy spectral template
gives an $R\/$-band polarization of 2--3\% for the remaining light.

Measurements of H$\alpha\/$ reported in Tables~2 and 3 are uncertain because
the redshift places the line center at the position of the O$_2$
A-band absorption.
Despite this unfortunate coincidence,
2M105144 earns its classification of Type~1.9 because H$\alpha\/$ possesses
an
$\sim$18,000~\kms\ (full width at zero intensity) component.
This broad feature is also detected in the polarized flux spectrum,
indicating that some of the flux from the inner BLR is scattered
into our line of sight.
No other features are identified in the polarized spectrum.

\subsubsection{2MASSI J130005.3+163214}

\citet{schmidt02} present and discuss the MMT observations of 
the Type~2 QSO 2M130005 that are redisplayed in Figure~3.
The dominant feature in the polarized spectrum of 2M130005, as
for 2M105144, is very broad ($\sim$18,000~\kms\ FWHM) H$\alpha$.
The S/N of the data is much higher
than for 2M105144, and
\citet{schmidt02} are able to deduce the polarization
of the continuum, broad H$\alpha$, and the narrow emission lines. 
In comparison to other QSOs in the sample, 2M130005 has by far
the reddest total flux (after subtracting the substantial
contribution of the host galaxy) and polarized flux spectra.
Not all of the \ion{Na}{1}~D feature is stellar in origin, since
it can be seen in $q' \times F_\lambda\/$.

In addition to the red polarized continuum, the Balmer decrement is
possibly very large since
H$\beta\/$ is not definitively detected in polarized flux.
However,
assuming that H$\beta\/$ is as broad as H$\alpha\/$ in polarized flux,
the extreme width of the line hinders its identification.
A rough limit on the strength of a possible broad H$\beta\/$
feature was estimated by
fitting a power-law to the polarized continuum redward of 5900~\AA\ and
avoiding H$\alpha\/$.
Subtraction of this fit from $q' \times F_\lambda\/$ reveals emission
from $\sim$5700~\AA\ to the blue edge of the spectrum at 4200~\AA\ that
could be the broad, blended lines of H$\beta\/$, H$\gamma\/$, and
higher-order Balmer lines.
The possible emission excess may indicate that the H$\beta\/$ flux
is as much as $0.14 \times 10^{-14}$~erg~cm$^{-2}$~s$^{-1}$, implying
a polarized Balmer decrement of only $\sim$3.

\subsubsection{2MASSI J171559.7+280717}

This object is the most distant of the Type~1.8--2s and one of the 
most luminous QSOs in the sample ($M_{K_s} = -28.1$).
Its redshift of $z = 0.524$ has shifted H$\alpha\/$ out of the
observed spectral range.
Weak, broad H$\beta\/$ is detected and we therefore classify 
2M171559 as Type~1.8.

The object is extremely faint ($V \sim 21.4$), but the spectropolarimetry
confirms that 2M171559 does indeed join 2M010835 in showing
a broadband polarization $>$3\%.
\citet{marble03} find significant extended flux around the AGN in the
WFPC2 F814W image.
Using this observation to estimate the contribution of starlight in
the spectropolarimetry suggests that about half of the light at
a rest frame wavelength of 5500~\AA\ is from the host galaxy.
This choice of host galaxy is quite uncertain as there are no
discernible stellar features superimposed on the faint continuum.
A host galaxy contribution this large implies an intrinsic polarization
within the $R\/$-band of over 15\% for the AGN. 

There is no evidence of the strong, narrow emission lines in the polarized
flux spectrum.
Close inspection of the spectrum of $q'\/$ reveals decreases in the 
polarization at [\ion{O}{3}]$\lambda\lambda$4959,5007,
[\ion{O}{2}]$\lambda$3727, and the narrow component of H$\beta\/$.
Some of the increase in polarized flux between
the location of narrow H$\beta\/$ and the [\ion{O}{3}] lines could
be due to polarized broad H$\beta\/$.
The S/N is insufficient to unambiguously identify broad H$\beta\/$
in polarized light and we have not attempted to measure the width
and strength of this possible feature.
Although the measurements are uncertain, the polarized
flux continuum (ignoring the region around H$\beta\/$) is much
redder than that of the host galaxy-subtracted total flux.

\section{Discussion}

\subsection{Spectral Properties}

Although nearly the full range of AGN spectral types is represented 
by the 21 objects described in the previous section, they may
not be representative of the 2MASS AGN sample as a whole.
The objects chosen for observation are primarily highly polarized.
This bias translates into a sample that is generally more luminous
in the near-IR and has higher near-IR--to--optical
flux ratios than a ``typical'' 2MASS QSO \citep{smith02}.
Caution should therefore be exercised in extrapolating the spectroscopic 
results for the highly polarized objects to the entire
sample.
The 2MASS sample is generally compared to the 
low-redshift ($z < 0.6$) members of the Palomar-Green (PG) QSO sample 
\citep{mschmidt83} in the following discussion.
The choice of the PG sample to represent optically-selected QSOs
is primarily dictated by the large amount of data available in the literature
for these objects.

\subsubsection{[\ion{O}{3}] Luminosity}

The 2MASS AGN discussed in the previous section have been classified
as QSOs based on the fact that they are as luminous at
2.2~$\mu$m as QSOs in other samples, not because of their observed
optical luminosity or color.
Indeed, nearly half of the spectropolarimetry sample shows evidence
that starlight from the host galaxy contributes more than 50\% of
the continuum light observed in the 1\arcsec --3\arcsec\ apertures 
employed.
The red colors and high fraction of optical host galaxy starlight
are consistent with
the nuclear regions being obscured and reddened by 
dust in our line of sight, resulting in a QSO sample that is apparently
optically underluminous and redder than previous cataloged QSOs.
The high polarizations observed
in the 2MASS sample
are also consistent with a sample of objects partially 
hidden from direct view by dust \citep{smith02}.
Since the dust extinction at $K\/$ is much less than that
experienced at optical
wavelengths, it stands to reason that a near-IR search would uncover
QSOs missed by surveys that rely on optical color
and luminosity selection criteria, and that the intrinsic luminosity
of these AGN would be more accurately estimated from a near-IR brightness
than by optical magnitude.

Data at longer wavelengths offer some support for the assumption that the
absolute $K_s\/$ magnitude, $M_{K_s}\/$, can be used to distinguish
QSOs from AGN of lower luminosity \citep{smith02}. 
Of the objects detected at 60~$\mu$m and 1.4~GHz, there is no large
systematic difference between 2MASS QSOs and optically-selected QSOs
for a given 2.2~$\mu$m luminosity. 
The validity of using $M_{K_s}\/$ as a measure
of intrinsic luminosity can also be verified optically by
inspecting the [\ion{O}{3}]$\lambda$5007 luminosities.
In the context of current ideas on the role that viewing orientation
plays in determining the observed properties of AGN
\citep[see e.g.,][]{antonucci93},
emission from the kpc-sized NLR is generally thought to be
more isotropic than the emission from the BLR and ionizing continuum,
although there is evidence for
anisotropy in at least
the higher ionization narrow emission lines in some AGN 
\citep[see e.g.,][]{jackson90,hes93,baker97}.
Despite these examples, the [\ion{O}{3}]$\lambda$5007 flux should,
to first order,
scale with the luminosity
of the ionizing nuclear continuum.

Figure~5 compares the [\ion{O}{3}]$\lambda$5007
luminosity of the 2MASS spectropolarimetry
sample with low-redshift PG QSOs.
Following \citet{boroson92}, the [\ion{O}{3}]$\lambda$5007
luminosity is given 
by 
\begin{displaymath}
M_{\rm [O~III]} = M_V - 2.5 \log ({\rm EW}_{\rm [O~III]}),
\end{displaymath}
where $M_V\/$ is calculated from the rest frame 5500~\AA\
flux density.
For both samples of objects,
a clear trend of increasing
[\ion{O}{3}]$\lambda$5007 luminosity with increasing $M_{K_s}\/$ is seen.
In addition, the strength of
the [\ion{O}{3}]$\lambda$5007 emission from the 2MASS QSOs is
within the large range exhibited by the optically-selected QSOs.
For the predominantly highly polarized 2MASS sample, the Type~1.5,
1.8 (2M171559), and 1.9 (2M010835) objects cannot be 
distinguished from the low-polarization, Type~1 PG QSOs.
The three Type~2 QSOs and 2M105144 are clustered around $M_{K_s} \sim -26$,
$M_{\rm [O~III]} \sim -24.5$; consistent with PG QSOs in having
fainter [\ion{O}{3}] emission for this lower near-IR brightness.

It is true that, in comparison with the PG sample,
the Type~1 2MASS QSOs typically have a lower
[\ion{O}{3}] luminosity for a given brightness at $K_s\/$.
Only seven of the eight Type~1 QSOs are included in
Figure~5 since [\ion{O}{3}]$\lambda$5007 is not detected in 2M125807. 
Such weak [\ion{O}{3}] is also seen in some PG QSOs, as four of 
74 objects in the \citet{boroson92} study do not show
emission from the NLR, and there are some PG QSOs with [\ion{O}{3}]
strengths as low as the 2MASS objects.
However, it is striking that the majority of PG QSOs show stronger
[\ion{O}{3}] emission than the highly polarized
Type~1 2MASS objects for a given
near-IR luminosity.

A proper comparison of the [\ion{O}{3}] emission between the near-IR
and optically-selected samples awaits measurements for a larger sample
of Type~1
2MASS objects, and the small number of objects presented here is
strongly biased toward those showing high optical polarization.
With these qualifications in mind, it appears that 2MASS is adept at
finding red QSOs that, although exhibiting broad permitted emission
lines in their total flux spectra, are underluminous in [\ion{O}{3}].
A possible explanation for a link between the adopted red near-IR color
selection 
and weak [\ion{O}{3}] strength may be that 2MASS finds 
AGN with a larger dust covering factor than is
typical for optically-selected QSOs.
\citet{borosonm92} suggest a similar situation for the low-ionization
broad absorption-line QSOs (BALQSOs)
within the sample of infrared-selected AGN discovered by the 
{\it Infrared Astronomical Satellite\/} \citep[{\sl IRAS\/};][]{low88}.

At the same time, there is no evidence from the objects of intermediate
spectral type in this small sample that suggests a {\it fundamental\/}
difference with UV-excess QSOs in terms of near-IR and [\ion{O}{3}]
luminosity.
For these AGN, it would appear that we have
a relatively unobstructed view of the NLR
that is powered by an
continuum source indistinguishable from an optically-selected 
QSO if viewed from the line-emitting region.

\subsubsection{Optical and Near-IR Continua}

Although dramatic differences are not seen between the 2MASS and PG
QSO samples in terms of [\ion{O}{3}] and near-IR luminosity,
clear trends are observed in the optical spectral 
index ($\beta_{\rm OPT}\/$).
Figure~6 plots both the near-IR spectral index ($\beta_{\rm IR}\/$;
determined from the 2MASS $JHK_s\/$ photometry) and the
$B - K_s\/$ color index against $\beta_{\rm OPT}\/$.
The optical continuum is much redder for the 2MASS objects than for
the comparison sample of PG QSOs with $z < 0.6$ measured by 
\citet{neugebauer87}. 
The displacement of the 2MASS QSOs away from the PG QSOs in Figure~6 is
generally consistent with that expected from reddening by dust, though
for consistency with the PG data, $\beta_{\rm OPT}\/$ has not been
corrected for the flux contribution 
made by the host galaxy. 
It is also the case that, for most of the 2MASS objects
discussed here, a significant
amount of the observed AGN flux is scattered into our line of sight.
The scattered light does not appear to experience the same amount of
reddening as the direct, unscattered light from the nucleus (see \S4.2), and 
dust scattering typically results in a much bluer scattered light spectrum.
Therefore, it is very difficult to ascribe a single reddening value
to account for the spectral energy distribution
throughout the entire optical/near-IR spectral region.

\subsubsection{Balmer Decrement}

Another reddening indicator is given by the
H$\alpha$/H$\beta\/$ flux ratio.
This quantity can be measured in 17 of the 21 objects observed.
Generally, H$\alpha$/H$\beta$ is much larger than typically found
for optically-selected QSOs, and in several objects it is measured to
be $>$10.
It can be seen in Figure~7 that, at least for Type~1 and 1.5 QSOs,
there is a trend between the Balmer decrement and the slope of optical
continuum
of the AGN.
As expected from dust extinction, larger Balmer decrements tend to
be observed for objects with redder optical continua.

For Figure~7, the estimated contributions from host galaxy starlight
have been subtracted from the spectrum.
The fluxes of the Balmer lines include both broad and narrow-line 
components, and [\ion{N}{2}]$\lambda\lambda$6548,6583 has not been
deblended from H$\alpha\/$.
The [\ion{N}{2}] lines are only significant relative to H$\alpha\/$
for the Types~1.9 and 2 objects, and even reducing the measured
H$\alpha\/$ flux by a factor of 2
to account for the blended forbidden lines still
implies a large Balmer decrement for these QSOs.

Figure~7 also includes a rough reference point for comparison of the 2MASS QSOs
with optically-selected QSOs.
The star in the figure represents the median high frequency power-law
fit to PG QSOs with $z < 0.6$ \citep{neugebauer87} and
H$\alpha$/H$\beta$ (=3.7) measured from the QSO template spectrum constructed
from the {\it Sloan Digital Sky Survey\/}
\citep[SDSS;][]{vandenberk01}.
Roughly half of the Type~1 2MASS QSOs and all of the Type~1.5-2 objects
have much larger Balmer decrements than the adopted optically-selected
composite QSO.

For each of the Type~1 and 1.5 objects that have their Balmer decrement
measured from the
$q' \times F_\lambda\/$ spectrum, H$\alpha\/$/H$\beta\/$
is smaller in polarized light than observed in the total flux spectrum
(Table~3).
However, a
proper comparison between the polarized and total flux Balmer decrements
requires that only the broad Balmer-line components be used to determine the
decrement since the polarized flux spectra of the 2MASS QSOs are generally
devoid of NLR features.
The narrow-line contribution to H$\alpha\/$ and H$\beta\/$
is not significant for the Type~1 objects, but by definition, the Type~1.5
QSOs exhibit Balmer lines with distinct broad-line and narrow-line components.
After deblending the components and subtracting an estimate of the
[\ion{N}{2}]$\lambda\lambda$6548,6583 flux from $F_{{\rm H}\alpha}\/$, a
broad-line Balmer decrement is derived.
The resolution of the observations does not permit direct measurement of
[\ion{N}{2}] and we have assumed that the forbidden-line flux equals
that of narrow H$\alpha\/$, close to the average line ratio for AGN
\citep[see e.g.,][]{veilleux87}.
It is clear from this exercise that broad H$\alpha\/$ dominates
the line flux in the five Type~1.5 QSOs that have measured Balmer decrements.
The broad-line Balmer decrements are found to be roughly the same as those
given in Table~3 for the total flux spectra, and therefore, the fact that
the decrement is smaller in polarized light is not caused by the inclusion
of narrow-line flux.

The smaller polarized Balmer decrements
suggest that
the scattered light may undergo less reddening than the total AGN light.
In fact, since $F_{{\rm H}\alpha}/F_{{\rm H}\beta} < 4$ for several objects,
either there is little
reddening from the nucleus to the scatterers and from the scatterers
to us, or the scattering efficiency is higher
at shorter wavelengths.
In the latter case, electron scattering would be ruled out.

\subsection{Polarization Properties}

Because  all of the highly polarized 2MASS QSOs
are included in this study,
these objects are largely responsible for the correlations
between the degree of polarization, near-IR luminosity,
and $B - K_s\/$ found by \citet{smith02}.
That is, AGN with high near-IR luminosity and near-IR--to--optical 
flux ratios tend to be highly polarized.
These trends are consistent with a model that consists of 1)~unpolarized,
reddened
AGN light that is observed directly, 2)~unpolarized starlight of the host
galaxy, and 3)~AGN light polarized by scattering into our line of sight.
Smith \etal argue that the majority of Type~1 objects are not highly
polarized because the unscattered nuclear light dominates any 
source of polarized light. 
From our analysis of selected objects, it appears clear that
the Type~2 QSOs show little broadband polarization because of 
host galaxy dilution, despite
the fact that their extremely red colors imply heavy extinction of 
nuclear light along our line of sight.
It may also be that the obscuration is so pervasive in the Type~2
QSOs that the light from the scattering region is also obscured.
The intermediate QSOs show the highest mean polarization.
For these, \citet{smith02} suggest that our direct
view of the nucleus is more heavily obscured than for a typical
Type~1 2MASS QSO, so that AGN light scattering into our view is
not swamped by direct nuclear light.
The fact that the QSOs of higher intrinsic luminosity tend to
be the highly polarized objects in the sample is a consequence
of the ability of the luminous AGN to better illuminate scattering
material.

\subsubsection{Continuum Polarization}

Starlight from the host galaxies of many 2MASS QSOs results in
continuum polarization being
substantially less than if
the AGN light could be observed in isolation. 
The 6000--7000~\AA\ polarizations for 14 QSOs
after the subtraction of the best estimates for the diluting
stellar continua are listed in Table~3.
The observed polarizations of the remaining objects are also
listed since these objects show no evidence of significant
``contamination'' by the host galaxy.
Not all objects exhibit high polarization even after the
stellar continuum is taken into account, but these results support
the assertion by \citet{smith02} that the distribution of
polarization of the 2MASS sample is even more 
distinct from optically-selected samples
than suggested from raw broadband polarimetry.
A striking example of the effects of dilution
is given by 2M130005
\citep[Figure~3 and][]{schmidt02}.
Both the broadband measurement and the subsequent 
spectropolarimetry yield $P < 3$\% in the $R\/$-band, but
correction for the obvious late-type stellar spectrum implies
an intrinsic polarization of the nuclear light of around 10\%.

A wide range of continuum slopes are observed for the polarized flux 
and the host
galaxy-subtracted total flux spectra.
The power-law spectral indices for both $F_\lambda\/$
and $q' \times F_\lambda\/$ are listed in Table~3.
It can be seen in Figure~8 that for most 2MASS QSOs
the polarized flux spectra are much bluer than observed for the optical
AGN total flux, but $\beta_{q' \times F_\lambda}\/$ is still generally redder
than the median 
optical continuum for the PG QSOs ($\beta_{\rm OPT} \sim -1.6$; Figure~7). 
Two factors are most likely responsible for bluer polarized flux
continua: the scattered light from the nuclear region experiences
less reddening than the light directly seen from the AGN, and/or
the scattering is more efficient in
the blue than at longer wavelengths.
The second effect, if applicable, rules out electron scattering
and strongly hints that the scattering particles are dust grains.
Although $\beta_{q' \times F_\lambda}\/$ tends to be bluer than
$\beta_{F_\lambda}\/$, the polarized flux can be very red.
Again, 2M130005 provides an extreme case of a red polarized 
continuum \citep{schmidt02}, and 2M170003 also shows
$\beta_{q' \times F_\lambda} > 2$.
The similarity of polarized and total flux spectral indices 
for each of these QSOs suggests that the light from the scattering
regions is reddened by an amount comparable to the total
flux spectrum.

Five objects have measured spectral indices for the polarized 
continuum that are redder than those for the total flux spectrum of the AGN.
Each of these measurements, however, is much more uncertain than
indicated by the formal error bars displayed in Figure~8.
The error bars do not fold in the uncertainty in the fraction
of host galaxy starlight.
Of the five QSOs,
polarizations $\lesssim$1.0\% are observed for 2M100121, 2M125807,
and 2M222221,
making measurement of the polarized continuum difficult.
Inspection of Figure~2 for 2M222221 does show that the polarized
continuum between H$\beta\/$ and H$\alpha$ is redder than 
$F_\lambda\/$ over the same wavelength range.
Low polarization also hampers measurement of $\beta_{q' \times F_\lambda}\/$
for 2M105144, 2M132917, and 2M222554; objects showing blue polarized
continua relative to their total flux spectra.
2M010835 and 2M171559 are the two remaining QSOs measured to have
redder polarized continua, but the value of $\beta_{F_\lambda}\/$
is highly dependent on the choice of AGN-to-host galaxy flux
ratio.
This is true for all objects with large fractions of starlight
in their spectra, mostly the Type~1.8--2 QSOs.
The fraction of host galaxy flux is highly uncertain for both
2M010835 and 2M171559 simply because of the weak continua observed
for these QSOs.

\subsubsection{Broad-Line Polarization and Hidden BLRs}

The vast majority of 2MASS QSOs included in this study exhibit
broad Balmer lines in their polarized flux spectra.
Some objects show polarization structure across the line profiles
(see below), but in general the line polarizations are similar to
that of the continuum in
degree and position angle.
This fact excludes synchrotron radiation in general as the source of 
polarization.
There are possibly only five QSOs
in the sample that do not have polarized broad 
permitted lines.
Three objects, 2M100121, 2M125807, and 2M222554 are very weakly
polarized, making the identification of spectral features
difficult.
The only Type~1 object that shows no evidence for polarized
emission lines is 2M125807, and it has the lowest polarization of the eight
objects observed.
The case of 2M222202 is detailed in \S3.3.6 and the available data
do not allow an unambiguous identification of weak polarized
H$\beta\/$ despite the high S/N polarized flux spectrum (Figures~2 and 4).
For this reason, a highly polarized synchrotron continuum cannot be
ruled out for 2M222202, but observations at three epochs 
spanning 2.5~yr have not shown the
object to be variable.
Emission from H$\beta\/$ also cannot be identified in the
much lower S/N polarized
flux spectrum of the optically faint QSO 2M171559. 

Given the resolution and S/N of the spectropolarimetry,
the polarized and total flux line widths of the Balmer lines
are roughly equivalent
for the Type~1--1.5 QSOs.
This implies that the scattering material is illuminated by an emission-line
spectrum that is not much different from the BLR spectrum observed
in our line of sight.
The Balmer line widths tabulated in Table~2 identify four
Type~1.5 QSOs with seemingly broader
Balmer lines in polarized flux than in total flux.
The decreased FWHM of the Balmer lines measured in the total flux spectrum
of 2M134915, 2M165939, and 2M222221 is
caused by the prominent narrow-line component largely absent from
the polarized line profile.
A Balmer emission-line component as broad as the polarized emission features
can be seen in the total flux spectrum for all three of these objects.
In the case of 2M222202, H$\beta\/$ is tentatively measured to be very
broad 
(FWHM $\sim$ 17,000~\kms ), although the emission feature is very
weak and may even be misidentified.

The Type 1.8--2 objects with measured Balmer-line widths for both
total and polarized flux spectra are represented by 2M010835, 2M105144,
and 2M130005.
The latter two objects were originally added to the sample to
search for hidden BLRs, and both, in fact, exhibit very broad
H$\alpha$ lines in polarized flux.
This broad-line component can also be seen in
the high S/N total flux spectrum of 2M105144 (Figure~3) and
has been noted in 2M130005 by \citet{schmidt02}.

2M130005 presents a dramatic example of a situation where 
emission from high-velocity clouds in the BLR is scattered into our
line of sight.
Though the broad H$\alpha\/$ line is extremely difficult to discern
in the total flux spectrum,
it appears as a very prominent, broad feature in polarized flux (Schmidt
\etal 2002).
Of course, hidden BLRs have been found in many
Seyfert~2s \citep{antonucci85,miller90,tran95}, 
NLRGs \citep{ogle97,cohen99},
and HIGs \citep{hinesw93,hines95,young96,goodrich96},
and these discoveries have greatly
promoted the idea that orientation of the nuclear region to the
line of sight largely determines the ``type'' of AGN that we
perceive.
The results from 2M105144 and 2M130005
suggest that near-IR--selected AGN will also provide a number
of similar objects; many with the intrinsic luminosity of a QSO.

Large polarized Balmer decrements and/or red polarized continua are
observed for 2M010607, 2M150113, 2M130005, 2M163700, and 2M170003, and
this may imply that the scattering region,
as well as the nucleus, is highly reddened along the line of sight,
or that light from the nucleus is reddened before reaching the scatterers.
It may be common for the scattering regions in 2MASS QSOs to
be reddened either by the dust torus that obscures the direct
view to the nucleus, or by dust in the body of the host galaxy.
A practical consequence of significant
extinction of the extended scattering regions 
for Type~2 objects is that hidden BLRs will be more difficult
to uncover polarimetrically.

Several 2MASS QSOs show higher polarization at H$\alpha\/$ and/or H$\beta\/$
compared to
the local continuum.
Good examples are 2M130005 and 2M134915,
and this property
can be attributed to dilution of the polarized light by an unpolarized
continuum.
Because the flux
of the emission lines is not as heavily diluted
as the polarized continuum, the 
polarization rises with the line profile.
The same effect is often seen in highly polarized
Seyfert~2s and radio galaxies, but
for many of these objects, careful subtraction of the stellar light
still leaves the broad emission lines more highly polarized than
the continuum \citep[see e.g.,][]{tran95,cohen99}.
Another unpolarized, featureless continuum (FC2) of unknown origin, is
generally invoked to bring the intrinsic continuum polarization to parity
with that of the BLR, avoiding the physically
untenable situation of having nuclear light, originating from a
very compact region, be less polarized than the light from
a more extended region surrounding the continuum source.
Within the S/N of the spectropolarimetry of this optically faint sample,
division of $q' \times F_\lambda\/$ by the total flux spectra after
subtraction of the estimated stellar continua yields
polarization levels at H$\alpha\/$ and H$\beta\/$ consistent with 
the local continuum polarization.
Therefore, the existence of FC2 in the 2MASS sample is not supported by the 
observations.

Beyond the effect of the host galaxy continuum on the polarization
of the broad Balmer lines relative to the continuum polarization, four
2MASS QSOs exhibit variations in polarization
across the Balmer-line profiles.
2M004118 and 2M151653 most clearly show changes in both
$P\/$ and $\theta\/$ at the positions of the BLR features
\citep[Figure~1 and][]{smith00}. 
In both cases, the blue wings of the lines are less polarized
than the red wings, but inspection of Figure~1 reveals that the
behavior of $\theta\/$ across H$\alpha\/$ is different for the two
QSOs.
For 2M132917 (Figure~1), deviations of $P\/$ and
$\theta\/$ from the level of the local continuum
appear to be restricted to the 
line core of H$\alpha\/$.
2M222221 (Figure~2) also shows complex rotations of $\theta\/$
across H$\alpha\/$ and, in contrast to 2M004118 and 2M151653,
diminished polarization in the line's red
wing.

The wavelength dependence of the polarization across the BLR emission
features implies a degree of complexity for the BLR and/or
scattering regions.
It also implies that the scattering material is located quite 
close to the BLR for these 2MASS QSOs, since polarization structure
is very difficult to explain if the BLR is unresolved from the 
vantage point of the scattering clouds.
Spectropolarimetry of several other Type~1 AGN have shown similar features.
\citet{goodrich94}
and \citet{smith97} have
identified cases among Seyfert~1 nuclei
\citep[see also][]{martel97,smithj02}, and \citet{cohen99}
have done the same from a small sample of broad-line radio galaxies
(BLRGs; namely, 3C~227 and 3C~445).
Apparently, the phenomenon is independent of the radio power of
the AGN.

\citet{cohen99} suggest a scenario to explain the complex
polarization position angle rotations observed in broad H$\alpha\/$
for 3C~445 using the assumption that the gas motion within the BLR
is not chaotic.
For instance, the BLR emission could be scattered off of the
inner wall of a dusty torus that is coplanar with the orbiting
BLR clouds.
If the emission line is broadened by the orbital motion of
the BLR gas, and not the thermal motion of the gas nor the motion
of the scatterers, the position angle difference between the red- and
blue-shifted line emission can be explained by the fact that
the two line components illuminate the scattering material from
different directions.
\citet{cohen99} point out that such a scenario works over a
restricted range of inclination angles.
The BLR also needs to be viewed directly
since broad lines are prominent in the total flux spectrum, but if
the inclination of the dust torus is too high, the scattering region
that gives rise to the wavelength dependence of $\theta\/$ across the
line will be obscured by the torus.
A similar scenario with narrower spectral features than 3C~445
could be applicable to 2M004118 and 2M222221, but the variety of 
wavelength dependences in $P\/$ and $\theta\/$ precludes a 
single geometry in all Type~1 objects.
For example, the Seyfert~1
Mrk~486 \citep{smith97} also requires at least two
polarized emission-line components to describe the polarization 
at H$\alpha\/$, but
instead of being shifted in wavelength, the components
have different widths.

\subsubsection{The Polarization of the NLR and the Location of the Scattering Material}

Emission from the NLR is observed to have low polarization in a wide variety
of polarized AGN samples where scattering of nuclear light into our line
of sight is thought to be the polarizing mechanism.
This is interpreted
as the result of the scattering material being located within, or
just exterior to, the NLR.
In this respect, the highly polarized 2MASS QSOs resemble
for example, 
polarized Seyferts, BLRGs, and HIGs
since they all  
show much reduced polarization in their
prominent narrow emission lines.
Close proximity of the scatterers to the nuclear region is also
required for the QSOs that show polarization
variations with wavelength
across their broad emission-line profiles (\S4.2.2).
Besides giving an indication of the location of the scattering regions,
the differentiation of the NLR polarization from the 
continuum and BLR validates the assumption made in 
\S2 that Galactic ISP is not the source of polarized flux for the
sample.

Polarized [\ion{O}{3}]$\lambda$5007 is convincingly detected
in six of
the 2MASS QSOs: 2M010835, 2M091848, 2M130005, 2M134915, 2M222202,
and 2M222221, plus 2M135852.
The level of polarization for the NLR is $<$3\% for all of these QSOs 
(last column of Table~3).
All spectral classes are represented in the small number of 
objects showing polarized [\ion{O}{3}]$\lambda$5007.
In general, the NLR shares a common polarization position angle with
the continuum and BLR, although the polarization of the NLR in 2M222202
is deduced from the rotation in $\theta\/$ seen in the
prominent narrow emission-lines.
The level of polarization of the NLR is low enough that transmission
through aligned dust grains within the body of the host galaxy
cannot be ruled out as the polarizing mechanism, but the common
polarization position angle of the NLR and the continuum observed
in most objects suggests that the material scattering the continuum
and BLR light also scatters and polarizes the inner NLR.
2M130005 provides evidence suggestive of the scattering material
being located throughout the NLR \citep{schmidt02}.
A polarization of $\sim$1.5\% is found for [\ion{O}{3}]$\lambda$5007,
whereas the lower ionization [\ion{S}{2}]$\lambda\lambda$6716,6731
lines are unpolarized.
In addition, the relative strengths of narrow H$\alpha\/$ and 
[\ion{N}{2}]$\lambda\lambda$6548,6583 are noticeably different in
polarized light {\it vs.\/} total flux, with H$\alpha\/$
being stronger the polarized light.
These observations are consistent with the regions producing the
[\ion{O}{3}] and narrow H$\alpha\/$ lines being located closer to
the nucleus, and with the scattering material being possibly intermixed
with the unpolarized NLR gas.
Similar evidence for such a stratified NLR is also seen in 
the HIG {\sl IRAS}~P09104+4109 \citep{hines99,tran00}.

If the continuum is polarized by scattering in 2M222202
(\S3.3.6), then the scatterers must be located closer to the 
nucleus than in the other 2MASS QSOs, since the polarized
spectrum is nearly featureless.
The scattering medium in this case may be electrons, given
that the scatterers are likely to be intermixed with the BLR gas.
Scattering by dust further out in the NLR may be responsible 
for the observed polarization in the rest of the sample.
Dust would naturally explain the diminished Balmer decrements and
blue continua of the polarized flux spectra, although electron
scattering coupled with reduced reddening along the line of sight
to the scattering region is also
consistent with the observations.
Ultraviolet spectropolarimetry may hold the key to identifying 
the scattering material in 2MASS QSOs since the shape of the spectrum
of UV polarized flux has been used to identify dust as the scatterers
in other reddened AGN \citep[e.g.,][]{hines01}.

\centerline{\ \ \ \ \ \ \ \ }

\subsection{Host Galaxies}

Direct imaging with {\sl HST\/} of 29 2MASS QSOs by \citet{marble03} 
shows that the sample exhibits a large range of AGN--to--host galaxy
flux ratios.
In addition, \citet{marble03} find that a wide variety 
of host galaxy morphologies and luminosities are represented, and
that no clear differences are seen between the 2MASS QSO hosts
and the galaxies hosting UV-excess QSOs.
Although there is little overlap between the 
samples, the spectropolarimetry also finds objects that show little
or no evidence for a significant contribution to their spectra from
starlight, as well as objects where the host galaxy dominates the
observed continuum.
Type~1.8--2 QSOs generally show the largest host galaxy-to-total
flux ratios in this sample, while stellar absorption features cannot
be definitively identified for most of the eight Type~1 objects.
These results are consistent with the suggestion by \citet{smith02}
that, unlike optically-selected QSO samples, dilution of the polarized
flux by starlight has a major effect
on the observed distribution of broadband polarization in the 2MASS QSO
sample (see \S4.2.1).

\centerline{\ \ \ \ \ \ \ \ }

\section{Conclusions and Summary}

Spectropolarimetry of 21 highly polarized and
narrow-line QSOs discovered by 2MASS reveals several important
aspects of this near-IR color-selected sample:

1. The observations are consistent
with scattering of nuclear continuum and emission from the BLR
by dust or electrons located exterior to the BLR as
the polarizing mechanism.
In general, the low polarization of the NLR implies that the 
scatterers cannot be situated much further from the ionizing
continuum source than the extent of the NLR.
The close proximity of the scattering material to the BLR in some objects is
also indicated by the observation of wavelength-dependent polarization
across their broad emission-line profiles.
For one object, 2M222202, the scatterers must be located very 
close to the nucleus because the broad emission lines
are essentially unpolarized.

2. Nearly all of the
polarized 2MASS QSOs
have broad Balmer emission lines detected in their polarized flux spectra. 

3. In addition to displaying a wide array of AGN spectral types, the
2MASS QSOs exhibit a huge range of optical continuum slopes 
and Balmer decrements.
These properties are consistent with various amounts of
reddening and extinction
of AGN light for these objects.
In general, the polarized flux continua are bluer
than the total flux spectra
(i.e., $P\/$ increases to the blue),
and the broad-line Balmer decrement is smaller in
the polarized light.
This trend can be explained if 
the scattering efficiency
increases with decreasing wavelength (small dust grain
scattering), or if the particles are electrons and the
scattered light experiences less reddening than the direct light from
the nucleus.
Although the polarized flux spectrum tends to be bluer than the total 
flux spectrum for these QSOs,
the polarized continuum can be extremely red and
large Balmer decrements can be measured for the polarized flux
spectrum.
For several objects, these properties indicate that there
is a significant amount of reddening of the scattering regions
situated around the nucleus.

4. The 2MASS sample includes a few objects with hidden BLRs.
Spectropolarimetry of the Type~2 QSO 2M130005 reveals an extremely 
broad H$\alpha\/$ emission line in polarized flux \citep{schmidt02}.
A similarly broad feature is also detected in both the
polarized and total flux spectra of 2M105144.

5. At optical wavelengths, $\gtrsim$50\% of the observed continuum for
many objects is from stars in the host galaxy.
The largest host galaxy contributions to the optical flux are 
generally found for the Type~1.8--2 QSOs.
This result is consistent with the finding of \citet{marble03}
from {\sl HST\/} imaging, that the fraction of host galaxy-to-AGN
light is generally larger for Type~1.8--2 objects than for Type~1 and 1.5
QSOs, and implies that the nuclear regions of Type~1.8--2 2MASS QSOs are 
typically more highly obscured in our line of sight.
The spectropolarimetry also supports the inference made by
\citet{smith02} that the low level
of broadband polarization observed for Type~1.8--2 QSOs is primarily due
to a large amount of unpolarized light included in the observation
aperture.
In particular, correction for the host galaxy starlight yields intrinsic 
optical polarizations of the nuclear continuum for some Type~1.8--2 QSOs of
$\sim$10--20\%.

6. The sample of eight Type~1 2MASS QSOs observed have
[\ion{O}{3}]$\lambda$5007 luminosities systematically lower
than PG QSOs of similar $M_{K_s}\/$.
Since the sample is biased toward highly
polarized objects, this may not be a property of the 2MASS QSOs
as a whole.
However, at least for the selected QSOs,
this finding suggests that dust obscuration extends over a larger 
solid angle than is typically inferred to account for the
relative numbers of optical spectral types in
simple unification schemes \citep[e.g.,][]{antonucci93}.
No difference in [\ion{O}{3}] luminosity is seen between
presumably more heavily obscured Type 1.5--1.9 2MASS 
QSOs and unobscured optically-selected QSOs.

It appears that the same basic model of an AGN surrounded by a dusty torus,
with
scattering material located above and below the toroidal plane,
can be applied to
Seyfert~1 and 2 nuclei, narrow- and broad-line radio galaxies, HIGs,
and now a large sample of
red AGN discovered by 2MASS.
Upon close examination, the 2MASS objects do not appear to differ 
systematically 
from previous optically-selected AGN in a variety of fundamental parameters.
The primary difference seems to be the amount of
extinction along our line of sight
to the nucleus.
Optical spectropolarimetry and [\ion{O}{3}] luminosities suggest
that the obscuring material does not cover all of the sky as seen by
the central engine, and that many 2MASS QSOs would appear as
typical UV/optical QSOs if viewed from a different perspective.

In retrospect, it may not be surprising that many
AGN are obscured to some degree from our direct view, since the nuclei are
often hosted by galaxies that contain large amounts of dust.
This situation was foreseen in Seyfert galaxies by \citet{rowan77},
but the magnitude of the nuclear obscuration
was not appreciated until the first IR surveys for AGN.
An exciting and challenging aspect of the new near-IR sample 
is that it represents a large, possibly dominant,
low-redshift AGN population \citep{cutri01}.
It is clear from even the small number of
objects observed in this study that the new search techniques are exploring
larger ranges of inclination angle and obscuration of the active
nucleus and its
immediate environment than have been probed by traditional surveys.

While the 2MASS QSOs so far identified add to our understanding of AGN
in the local universe, it is likely that more highly obscured
objects are excluded by the selection criteria.
At sufficiently high column depths, the $J\/$-band flux from
the nucleus will fall below the 2MASS sensitivity limit or below that of the
host galaxy, making color selection inefficient.
Surveys at longer wavelengths would alleviate this problem.
The {\it Space Infrared Telescope Facility\/} ({\sl SIRTF\/})
has the potential to uncover many more obscured AGN, but its coverage 
of the sky will be limited.

\acknowledgments

We thank the
National Aeronautics and Space Administration (NASA)
and the Jet Propulsion Laboratory (JPL) for 
support through {\sl SIRTF\/}/MIPS and Science Working Group contracts 960785
and 959969 to The University of Arizona.
Polarimetric instrumentation at Steward Observatory is maintained, in part,
through support by National Science Foundation (NSF) grants AST 97--30792
and AST 98--03072.
We also thank an anonymous referee for suggestions that improved the
manuscript.
This publication makes use of data products from the Two-Micron All Sky Survey, which is a joint project of the University of Massachusetts
and the Infrared Processing and Analysis
Center/California Institute of Technology,
funded by NASA
and the NSF.


\clearpage

\begin{deluxetable}{lrcrrrlccrrr}
\tabletypesize{\scriptsize}
\tablecolumns{12}
\tablewidth{0pc}
\rotate
\tablecaption{Objects and Observations}
\tablehead{
\colhead{Object}  &
\colhead{$z\/$} & \colhead{Type} &
\colhead{$M_{K_s}\/$\tablenotemark{a}} &
\colhead{$K_s\/$\tablenotemark{b}} &
\colhead{$B-K_s\/$\tablenotemark{b}} &
\colhead{UT Date} &
\colhead{Tel.} & Slit Width&
\colhead{Exp.} &
\colhead{$P\/$\tablenotemark{c}} &
\colhead{$\theta\/$\tablenotemark{c}} \\
\colhead{(2MASSI J)} & \colhead{} & \colhead{} & \colhead{} & \colhead{} &
\colhead{} & 
\colhead{} & \colhead{} & \colhead{(\arcsec )} &
 \colhead{(s)} & \colhead{(\%)} &
 \colhead{(\dgr )}}
\startdata

004118.7+281640 & 0.194 & 1 & $-$27.29 & 12.50 & 3.40 & 1999 Oct 14 & Bok & 3.0 & 7200 & 2.18$\pm$0.02 & 97.9$\pm$0.3 \\
 & & & & & & 2002 Jul 6 & MMT & 1.1 & 1920 & 2.28$\pm$0.02 & 95.2$\pm$0.3 \\
 & & & & & & ave. &   &   & 9120 & 2.23$\pm$0.01 & 96.9$\pm$0.2 \\
010607.7+260334 & 0.411 & 1 & $-$27.58 & 14.61 & $>$6.4 & 2000 Jan 9, 10 & Bok & 2.0,3.0 & 19200 & 7.84$\pm$0.14 & 118.4$\pm$0.5 \\
010835.1+214818 & 0.285 & 1.9 & $-$27.64 & 13.46 & 6.54 & 1999 Oct 13, 15 & Bok & 2.0 & 22400& 5.07$\pm$0.11 & 118.9$\pm$0.6 \\
091848.6+211717 & 0.149 & 1.5 & $-$26.65 & 12.55 & 5.95 & 2000 Jan 9, 10 & Bok & 3.0 & 24800 & 6.49$\pm$0.02 & 153.4$\pm$0.1 \\
100121.1+215011 & 0.248 & 2 & $-$25.79 & 14.68 & 5.52 & 2002 Feb 15 & MMT & 1.1 & 3840 & 0.73$\pm$0.16 & 150.5$\pm$6.0 \\
105144.2+353930 & 0.158  & 1.9 & $-$25.77 & 13.54 & 5.06 & 2002 Feb 15 & MMT & 1.5 & 3840 & 1.18$\pm$0.11 & 30.8$\pm$2.7 \\
125807.4+232921 & 0.259 & 1 & $-$27.09 & 13.45 & 3.85 & 2000 Jan 10 & Bok & 3.0 & 4800 & 1.00$\pm$0.03 & 107.7$\pm$1.0 \\
130005.3+163214\tablenotemark{d} & 0.080 & 2 & $-$25.84 & 11.86 & 5.24 & 2001 Apr 1 & MMT & 1.1 & 3840 & 2.58$\pm$0.03 & 50.8$\pm$0.3 \\
 & & & & & & 2002 Jan 5 & Bok & 3.0 & 2400 & 1.81$\pm$0.77 & 46.5$\pm$1.2 \\
 & & & & & & 2002 Jul 4, 5 & MMT & 1.1 & 8640 & 2.51$\pm$0.01 & 43.7$\pm$0.2 \\
 & & & & & & ave. & & & 14880 & 2.76$\pm$0.01 & 45.3$\pm$0.1 \\
132917.5+121340 & 0.203 & 1 & $-$25.78 & 14.12 & 4.58 & 2001 Mar 31 & MMT & 1.1 & 3840 & 1.37$\pm$0.05 & 13.8$\pm$1.0 \\
134915.2+220032 & 0.062 & 1.5 & $-$24.87 & 12.24 & 3.27 & 2000 May 7 & Bok & 3.0 & 9600 & 1.84$\pm$0.03 & 108.2$\pm$0.5 \\
 & & & & & & 2001 Mar 30 & MMT & 1.5 & 8640 & 1.64$\pm$0.03 & 104.0$\pm$0.5 \\
 & & & & & & ave. & & & 18240 & 1.78$\pm$0.02 & 106.9$\pm$0.3 \\
135852.5+295413\tablenotemark{e} & 0.113 & 1 & $-$25.58 & 12.85 & 4.15 & 2000 May 7 & Bok & 3.0 & 12800 & 4.59$\pm$0.02 & 23.0$\pm$0.1 \\
150113.1+232908 & 0.258 & 1 & $-$27.24 & 13.46 & 5.84 & 2001 Mar 30, Apr 1 & MMT & 1.5,1.1 & 12480 & 3.40$\pm$0.04 & 156.0$\pm$0.3 \\
151653.2+190048\tablenotemark{f} & 0.190 & 1 & $-$28.35 & 11.41 & 4.39 & 2000 May 9 & Bok & 3.0 & 6720 & 9.27$\pm$0.01 & 106.5$\pm$0.1 \\
163700.2+222114 & 0.211 & 1.5 & $-$26.44 & 13.59 & 5.41 & 2002 Jul 5 & MMT & 1.1 & 7200 & 2.49$\pm$0.04 & 116.4$\pm$0.4 \\
165939.7+183436\tablenotemark{g} & 0.170 & 1.5 & $-$26.59 & 12.91 & 5.29 & 1999 Oct 13, 14, 15 & Bok & 3.0,2.0 & 9600 & 5.34$\pm$0.06 & 158.4$\pm$0.3 \\
 & & & & & & 2001 Mar 31 & MMT & 1.1 & 960 & 5.44$\pm$0.09 & 160.0$\pm$0.5 \\
 & & & & & & ave. & & & 10560 & 5.33$\pm$0.03 & 158.8$\pm$0.2 \\
170003.0+211823 & 0.596 & 1.5 & $-$28.30 & 14.88 & 7.21 & 2001 Mar 31 & MMT & 1.1 & 5760 & 10.96$\pm$0.15 & 108.4$\pm$0.4 \\
 & & & & & & 2002 Jul 4 & MMT & 1.1 & 7200 & 11.10$\pm$0.08 & 108.4$\pm$0.2 \\
 & & & & & & ave. & & & 12960 & 11.06$\pm$0.04 & 108.7$\pm$0.1 \\
171559.7+280717 & 0.524 & 1.8 & $-$28.14 & 14.63 & $>$6.4 & 2002 Feb 19 & MMT & 1.1 & 1920 & 3.62$\pm$1.11 & 0.3$\pm$8.8 \\
 & & & & & & 2002 Jul 6 & MMT & 1.1 & 7200 & 5.03$\pm$0.18 & 1.4$\pm$1.0 \\
 & & & & & & ave. & & & 9120 & 5.16$\pm$0.08 & 1.5$\pm$0.4 \\
222202.2+195231 & 0.366 & 1.5 & $-$28.60 & 13.30 & 6.20 & 1999 Oct 13, 15 & Bok & 2.0 & 19200 & 11.41$\pm$0.11 & 118.1$\pm$0.3 \\
 & & & & & & 2002 Jan 6 & Bok & 2.0,3.0 & 5760 & 12.51$\pm$0.25 & 116.9$\pm$0.6 \\
 & & & & & & 2002 Jul 4 & MMT & 1.1 & 4800 & 11.59$\pm$0.03 & 118.8$\pm$0.1 \\
 & & & & & & ave. & & & 29760 & 11.04$\pm$0.04 & 118.8$\pm$0.1 \\
222221.1+195947 & 0.211 & 1.5 & $-$27.10 & 12.92 & 4.58 & 1999 Oct 15 & Bok & 3.0 & 6400 & 1.15$\pm$0.04 & 156.4$\pm$0.9 \\
 & & & & & & 2002 Jul 8 & MMT & 1.5 & 1920 & 0.96$\pm$0.03 & 165.2$\pm$0.8 \\
 & & & & & & ave. & & & 8320 & 1.02$\pm$0.02 & 161.7$\pm$0.6 \\
222554.2+195837 & 0.147 & 2 & $-$25.65 & 13.49 & 5.31 & 2002 Jul 5 & MMT & 1.1 & 7200 & 0.28$\pm$0.03 & 0.7$\pm$3.4 \\
230307.2+254503 & 0.331 & 1 & $-$26.70 & 14.50 & 6.21 & 1999 Oct 14, 16 & Bok & 2.0 & 22400 & 3.31$\pm$0.09 & 137.3$\pm$0.8 \\
 & & & & & & 2002 Jul 8 & MMT & 1.1 & 2400 & 4.66$\pm$0.19 & 138.3$\pm$1.2 \\
 & & & & & & ave. & & & 24800 & 3.47$\pm$0.04 & 138.2$\pm$0.3 \\
\enddata

\tablenotetext{a}{The K-corrected absolute $K_s\/$ magnitude is from 
Smith \etal (2002) and is based on 2MASS photometry.
The listed values assume that $H_0 = 75~$\kms~Mpc$^{-1}$, $q_0 = 0$, and
$\Lambda = 0$.}
\tablenotetext{b}{Data are from the 2MASS Point Source Catalog.}
\tablenotetext{c}{The flux-weighted mean optical linear
polarization in the 5000--8000~\AA\ band (observed frame).
The listed degree of linear polarization has not been
corrected for statistical bias.}
\tablenotetext{d}{Observations reported in \citet{schmidt02}.} 
\tablenotetext{e}{2M135852 is not considered a member of the 2MASS sample 
of QSOs since $J - K_s < 2$.}
\tablenotetext{f}{Observations reported in \citet{smith00}.} 
\tablenotetext{g}{Observations obtained with the Bok Telescope
reported in \citet{smith00}.} 

\end{deluxetable}

\clearpage

\begin{deluxetable}{lcccccccccc}
\tabletypesize{\scriptsize}
\tablecolumns{11}
\tablewidth{0pc}
\rotate
\tablecaption{Emission-Line Properties}
\tablehead{
\colhead{Object}  &
\colhead{EW$_{{\rm H}\beta}$\tablenotemark{a}} &
\colhead{EW$_{{\rm H}\alpha}$\tablenotemark{b}} &
\colhead{EW$_{\rm [O~III]}$} &
\colhead{FWHM$_{{\rm H}\beta}$\tablenotemark{a}} &
\colhead{FWHM$_{{\rm H}\alpha}$\tablenotemark{b}} &
\colhead{$F_{{\rm H}\beta}$\tablenotemark{a,{\rm c}}} &
\colhead{$F_{{\rm H}\alpha}$\tablenotemark{b,{\rm c}}} &
\colhead{$F_{\rm [O~III]}$\tablenotemark{c}} &
\colhead{Log $L_{\rm [O~III]}$\tablenotemark{d}} &
\colhead{Notes} \\
\colhead{(2MASSI J)} & 
\colhead{(\AA )} & 
\colhead{(\AA )} & 
\colhead{(\AA )} & 
\colhead{(\kms )} & 
\colhead{(\kms )} & 
\colhead{} & 
\colhead{} & 
\colhead{} & 
\colhead{(erg s$^{-1}$)} & 
\colhead{} \\
\colhead{} & 
\colhead{$F_\lambda$/$q' \times F_\lambda$} & 
\colhead{$F_\lambda$/$q' \times F_\lambda$} & 
\colhead{$F_\lambda$/$q' \times F_\lambda$} & 
\colhead{$F_\lambda$/$q' \times F_\lambda$} & 
\colhead{$F_\lambda$/$q' \times F_\lambda$} & 
\colhead{$F_\lambda$/$q' \times F_\lambda$} & 
\colhead{$F_\lambda$/$q' \times F_\lambda$} & 
\colhead{$F_\lambda$/$q' \times F_\lambda$} & 
\colhead{$F_\lambda$/$q' \times F_\lambda$} & 
\colhead{}}
\startdata
\sidehead{Type 1:}
004118.7+281640 & 81/\nodata & 500/130: & 8/\nodata & 2280/\nodata & 1960/3200: &
8.07/\nodata & 29.5/0.16: & 0.72/\nodata & 41.95/\nodata & 1,2 \\
010607.7+260334 & 71/87 & N/A & 20/\nodata & 1740/1880 & N/A &
0.20/0.02 & N/A & 0.06/\nodata & 41.71/\nodata & \\
125807.4+232921 & 25/\nodata & 145/\nodata & \nodata/\nodata & 2350/\nodata & 2010/\nodata &
1.09/\nodata & 4.30/\nodata & \nodata/\nodata & \nodata/\nodata & \\
132917.5+121340 & 75/180 & 300/\nodata & 29/$<$15 & 3660/4610 & 3610/\nodata &
1.18/0.05 & 4.75/\nodata & 0.45/$<$0.01 & 41.79/$<$39.7 & 1 \\
135852.5+295413 & 36/68 & 275/550 & 14/6 & 6720/6590 & 5440/5430 &
2.21/0.18 & 15.00/1.14 & 0.75/0.02 & 41.41/39.77 & \\
150113.1+232908 & 32/85 & 225/890 & 8/$<$7 & 2530/3170 & 2540/2580 &
0.23/0.03 & 2.71/0.18 & 0.06/$<$0.01 & 41.19/$<$39.8 & \\
151653.2+190048 & 87/92 & 645/440 & 16/\nodata & 4040/4840 & 3680/4100 &
18.47/1.95 & 110.00/6.35 & 3.36/\nodata & 42.60/\nodata & 1 \\
230307.2+254503 & 49/40 & 80:/\nodata & 10/\nodata & 2300/1550 & 1800:/\nodata &
0.22/0.01 & 0.6:/\nodata & 0.04/\nodata & 41.33/\nodata & 3 \\

\sidehead{Type 1.5:}
091848.6+211717 & 44/74 & 240/230 & 62/18 & 1980/2270 & 1790/1800 &
0.85/0.11 & 6.54/0.35 & 1.19/0.03 & 41.89/40.21 & 4 \\
134915.2+220032 & 63/260 & 310/480 & 185/73 & 1900/4580 & 1800/2620 &
2.69/0.17 & 15.90/0.49 & 7.44/0.05 & 41.82/39.66 & 5 \\
163700.2+222114 & 25/49 & 220/320 & 42/\nodata & 1270/1930 & 2910/3700 &
0.12/0.01 & 2.47/0.09 & 0.32/\nodata & 41.69/\nodata & \\
165939.7+183436 & 45/117 & 240/500 & 112/\nodata & 4110/15700 & 4270/7430 &
1.02/0.14 & 6.27/0.49 & 2.48/\nodata & 42.34/\nodata & 5,6 \\
170003.0+211823 & 123/90 & N/A & 64/\nodata & 2630/2980 & N/A &
0.35/0.03 & N/A & 0.19/\nodata & 42.74/\nodata & 6 \\
222202.2+195231 & 206/70: & N/A & 173/\nodata & 2040/17000: & N/A &
1.22/0.06: & N/A & 1.03/\nodata & 42.84/\nodata & 7 \\
222221.1+195947 & 118/240 & 880/1110 & 65/25 & 4260/7910 & 3390/6860 &
8.35/0.17 & 46.80/0.57 & 4.49/0.02 & 42.84/40.37 & 1 \\

\sidehead{Type 1.8--2:}
010835.1+214818 & 56/43 & 430/480: & 360/120 & 870/1320 & 1730/3100: &
0.27/0.01 & 2.21/0.16: & 2.43/0.04 & 42.91/41.10 & 3,5 \\
100121.1+215011 & 7/\nodata & 190/\nodata & 24/\nodata & 1090/\nodata & 1830/\nodata &
0.03/\nodata & 0.80/\nodata & 0.09/\nodata & 41.30/\nodata & 3,5 \\
105144.2+353930 & 5/\nodata & 270:/\nodata & 97/\nodata & 930/\nodata & 2000:/9000: &
0.04/\nodata & 2.09:/0.10: & 0.55/\nodata & 41.61/\nodata & 5,8 \\
130005.3+163214 & 5/\nodata & 57/415 & 57/56 & 1210/\nodata & 2060/18400 &
0.18/$<$0.14 & 3.67/0.45 & 1.61/0.03 & 41.40/39.61 & 5 \\
171559.7+280717 & 130/\nodata & N/A & 450:/\nodata & 1090/\nodata & N/A &
0.19/\nodata & N/A & 0.6:/\nodata & 43:/\nodata & 8 \\
222554.2+195837 & 4/\nodata & 72/\nodata & 43/\nodata & 960/\nodata & 2190/\nodata &
0.08/\nodata & 1.06/\nodata & 0.58/\nodata & 41.56/\nodata & 4,5 \\

\enddata

\tablenotetext{a}{The H$\beta\/$ equivalent width (EW$_{{\rm H}\beta}\/$), line width
(FWHM$_{{\rm H}\beta}\/$),
and flux ($F_{{\rm H}\beta}\/$)
include the contributions from both the narrow- and broad-line components.}
\tablenotetext{b}{Measurements of EW$_{{\rm H}\alpha}\/$,
FWHM$_{{\rm H}\alpha}\/$, and $F_{{\rm H}\alpha}\/$
include the contributions from [N~II]$\lambda\lambda$6548,6583 and
both the narrow- and broad-line components
of H$\alpha\/$.}
\tablenotetext{c}{Emission-line fluxes have units of 
10$^{-14}$~erg~cm$^{-2}$~s$^{-1}$.}
\tablenotetext{d}{Logarithm of the [O~III]$\lambda$5007
luminosity.}

\tablecomments{(1) Structure in $q'\/$ and $\theta\/$ in the line profile
of H$\alpha\/$. 
(2) Data corrected for Galactic ISP.
(3) H$\alpha\/$ located near the edge of the observed spectrum.
(4) The terrestrial A-band O$_2$ absorption band affects the red wing
of H$\alpha\/$ + [N~II].
(5) H$\alpha\/$ includes significant [N~II]$\lambda\lambda$6548,6583.
(6) The extreme blue portion of H$\alpha\/$ or H$\beta\/$ is affected
by O$_2$ A-band absorption.
(7) Identification of a polarized spectral feature with H$\beta\/$
is uncertain.
(8) The H$\alpha\/$ + [N~II] flux ([O~III]$\lambda$5007 for
2M171559) is uncertain because the redshifted lines fall within the
O$_2$ A-band absorption feature.
}

\end{deluxetable}

\clearpage

\begin{deluxetable}{lcrrccccccc}
\tabletypesize{\scriptsize}
\tablecolumns{11}
\tablewidth{0pc}
\rotate
\tablecaption{Continuum Properties, Line Ratios, and Polarizations}
\tablehead{
\colhead{Object}  &
\colhead{$F_{\rm gal}$/$F_{\rm Total}$\tablenotemark{a}} &
\colhead{$P\/$\tablenotemark{b}} &
\colhead{$\theta\/$\tablenotemark{b}} &
\colhead{$\beta\/$\tablenotemark{c}} &
\colhead{$F_{\rm [O~III]}$/$F_{{\rm H}\beta}$\tablenotemark{d}} &
\colhead{$F_{{\rm H}\alpha}$/$F_{{\rm H}\beta}$\tablenotemark{d}} &
\colhead{$F_{\rm Fe~II}$/$F_{{\rm H}\beta}$\tablenotemark{d}} &
\colhead{$P_{{\rm H}\beta}$\tablenotemark{e}} &
\colhead{$P_{{\rm H}\alpha}$\tablenotemark{e}} &
\colhead{$P_{\rm NLR}$\tablenotemark{f}} \\
\colhead{(2MASSI J)} & 
\colhead{} &
\colhead{(\%)} &
\colhead{(\dgr )} &
\colhead{} &
\colhead{} &
\colhead{} &
\colhead{} &
\colhead{(\%)} &
\colhead{(\%)} &
\colhead{(\%)} \\
\colhead{} &
\colhead{} &
\colhead{} &
\colhead{} &
\colhead{$F_\lambda$/$q' \times F_\lambda$} & 
\colhead{$F_\lambda$/$q' \times F_\lambda$} & 
\colhead{$F_\lambda$/$q' \times F_\lambda$} & 
\colhead{$F_\lambda$/$q' \times F_\lambda$} & 
\colhead{line/cont} & 
\colhead{line/cont} &
\colhead{}}
\startdata
\sidehead{Type 1:}
004118.7+281640 & \nodata & 2.35$\pm$0.02 & 96.3$\pm$0.2 & $-$2.2/$-$1.9 &
0.1/\nodata & 3.7/\nodata & 1.4/\nodata & \nodata/2.3 & 0.5:/2.1 & \nodata \\
010607.7+260334 & \nodata & 7.73$\pm$0.20 & 117.3$\pm$0.8 & +1.3/+0.5 &
0.3/\nodata & N/A & 0.8/\nodata & 9.1/8.6 & N/A & \nodata \\
125807.4+232921 & 0.1:\tablenotemark{\dag} & 1.05$\pm$0.06 & 111.4$\pm$1.7 & $-$2.2/$-$1.0: &
\nodata/\nodata & 3.9/\nodata & 2.1/\nodata & \nodata/2: & \nodata/3: & \nodata \\
132917.5+121340 & \nodata & 1.32$\pm$0.07 & 11.4$\pm$1.6 & $-$0.5/$-$2.0: &
0.4/$<$0.1 & 4.0/\nodata & 0.7/\nodata & 4.1/2: & \nodata/3: & $<$1 \\
135852.5+295413 & 0.57 & 9.10$\pm$0.07 & 23.4$\pm$0.2 & +0.7/$-$0.9 &
0.3/0.1 & 6.8/6.1 & 0.2/\nodata & 8.5/12.8 & 7.6/7.9 & 2.4 \\
150113.1+232908 & 0.59 & 8.25$\pm$0.14 & 157.0$\pm$0.5 & +2.3/+0.0 &
0.3/$<$0.1 & 11.6/7.1 & 0.7/\nodata & 10.7/12.2 & 6.5/6.1 & $<$3 \\
151653.2+190048 & \nodata & 9.24$\pm$0.02 & 107.0$\pm$0.1 & $-$0.9/$-$2.1 &
0.2/\nodata & 6.0/3.3 & 1.1/1.0 & 10.6/10.7 & 5.8/7.5 & \nodata \\
230307.2+254503 & \nodata & 2.93$\pm$0.05 & 138.8$\pm$0.5 & +0.9/$-$0.3 &
0.2/\nodata & 3:/\nodata & 3.3/\nodata & 3.7/7.0  & \nodata/5.0 & \nodata \\

\sidehead{Type 1.5:}
091848.6+211717 & 0.37 & 10.01$\pm$0.07 & 153.3$\pm$0.2 & +1.0/$-$0.6 &
1.4/0.2 & 7.7/3.3 & 1.1/1.4 & 12.4/12.7 & 5.4/8.0 & 2.1 \\
134915.2+220032 & 0.47 & 2.62$\pm$0.04 & 108.0$\pm$0.4 & +0.8/+0.0 &
2.8/0.3 & 5.9/2.9 & 0.8/\nodata & 6.2/4.7 & 3.1/3.7 & $<$1 \\
163700.2+222114 & 0.51 & 4.46$\pm$0.10 & 115.6$\pm$0.6 & +2.0/+0.9 &
1.6/\nodata & 12.6/9.2 & 1.1/\nodata & 8.3/7.4 & 3.6/5.4 & \nodata \\
165939.7+183436 & 0.26 &  7.14$\pm$0.12 & 158.8$\pm$0.5 & $-$0.3/$-$0.7 &
2.4/\nodata & 6.2/3.5 & \nodata/\nodata & 14.0/8.6 & 7.8/7.6 & \nodata \\
170003.0+211823 & \nodata & 11.40$\pm$0.05 & 107.9$\pm$0.1 & +2.1/+2.0 &
0.5/\nodata & N/A & 0.4/\nodata & 8.0/12.4 & N/A & \nodata \\
222202.2+195231 & \nodata & 10.40$\pm$0.05 & 118.9$\pm$0.1 & +0.6/$-$1.0 &
0.8/\nodata & N/A & 0.3/\nodata & 8:/14.1 & N/A & 0.8 \\
222221.1+195947 & 0.3:\tablenotemark{\dag} & 1.15$\pm$0.04 & 163.7$\pm$1.0 & $-$1.8/$-$0.2: &
0.5/0.1 & 5.6/3.4 & 0.2/\nodata & 2.0/2: & 1.2/3: & $<$1 \\

\sidehead{Type 1.8--2:}
010835.1+214818 & 0.8: & 22:$\pm$1: & 120.0$\pm$1.2 & $-$2.3:/$-$0.6 &
9.1/3.5 & 8.2/15: & \nodata/\nodata & 4.1/24: & 7:/40: & 1.6 \\
100121.1+215011 & 0.84 & 2.58$\pm$1.11 & 152.7$\pm$12.4 & $-$0.7/+0.5: &
3.3/\nodata & 31/\nodata & \nodata/\nodata & \nodata/20: & \nodata/30: & \nodata \\
105144.2+353930 & 0.71 & 2.72$\pm$0.55 & 38.3$\pm$5.8 & +2.4/+0.6: &
15/\nodata & 57:/\nodata & \nodata/\nodata & \nodata/20: & 5:/10: & \nodata \\
130005.3+163214 & 0.81 & 9.50$\pm$0.07 & 44.7$\pm$0.2 & +3.6/+3.0 &
8.9/\nodata & 20/$>$3 & \nodata/\nodata & \nodata/11.9 & 12.2/9.8 & 1.6 \\
171559.7+280717 & 0.5:\tablenotemark{\dag} & 18.5:$\pm$0.4: & 2.4$\pm$0.7 & $-$0.0:/+0.8: &
3:/\nodata & N/A & \nodata/\nodata & \nodata/17: & N/A & \nodata \\
222554.2+195837 & 0.80 & 1.00$\pm$0.24 & 7.2$\pm$6.8 & +1.2/+0.4: &
7.8/\nodata & 14/\nodata & \nodata/\nodata & \nodata/7: & \nodata/5: & \nodata \\

\enddata

\tablenotetext{a}{The adopted ratio of host galaxy starlight to total
light at a rest frame wavelength of 5500~\AA\ determined from
observed stellar absorption features or from \citet{marble03} ($^\dag$).}
\tablenotetext{b}{Flux-weighted linear polarization in the 
6000--7000~\AA\ band (observed frame).  Values of $P\/$
are corrected for dilution
by unpolarized starlight from the host galaxy using the estimate given
by $F_{\rm gal}$/$F_{\rm Total}\/$ in this table.  The degree of 
polarization is not corrected for statistical bias.}
\tablenotetext{c}{The continuum spectral slopes for the total flux 
($F_\lambda\/$) and polarized flux ($q' \times F_\lambda\/$) spectra.
Host galaxy starlight has been subtracted from $F_\lambda\/$
before fitting a power law to the continuum.}
\tablenotetext{d}{The H$\alpha\/$ and H$\beta\/$ fluxes include the
contributions
from both the
narrow- and broad-line components.  In the case of H$\alpha\/$, flux
from [N~II]$\lambda\lambda$6548,6583 is also included.}
\tablenotetext{e}{The first entry in the column is the estimate of the
emission-line polarization based on the measurements of the line
flux in the total and polarized flux spectra (Table~2).  The second entry is
the continuum polarization at the wavelength of the line determined
by the power-law fits to the total and polarized spectra (see text).}
\tablenotetext{f}{The polarization of the emission from the NLR, based
on measurements of [O~III]$\lambda$5007.}

\end{deluxetable}

\clearpage


\begin{figure}
\figurenum{1}
\vspace{8.6in}
\includegraphics{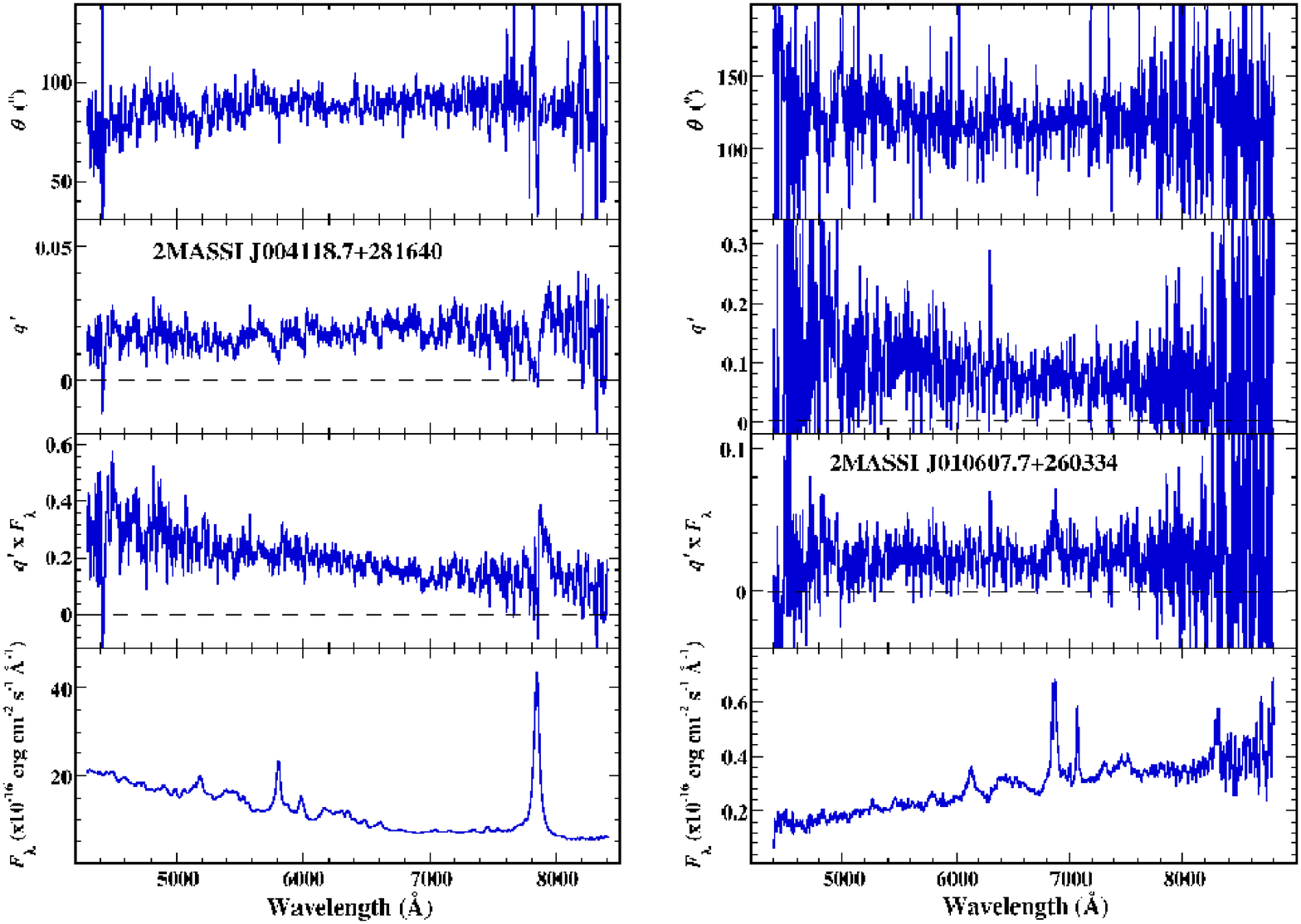}
\includegraphics{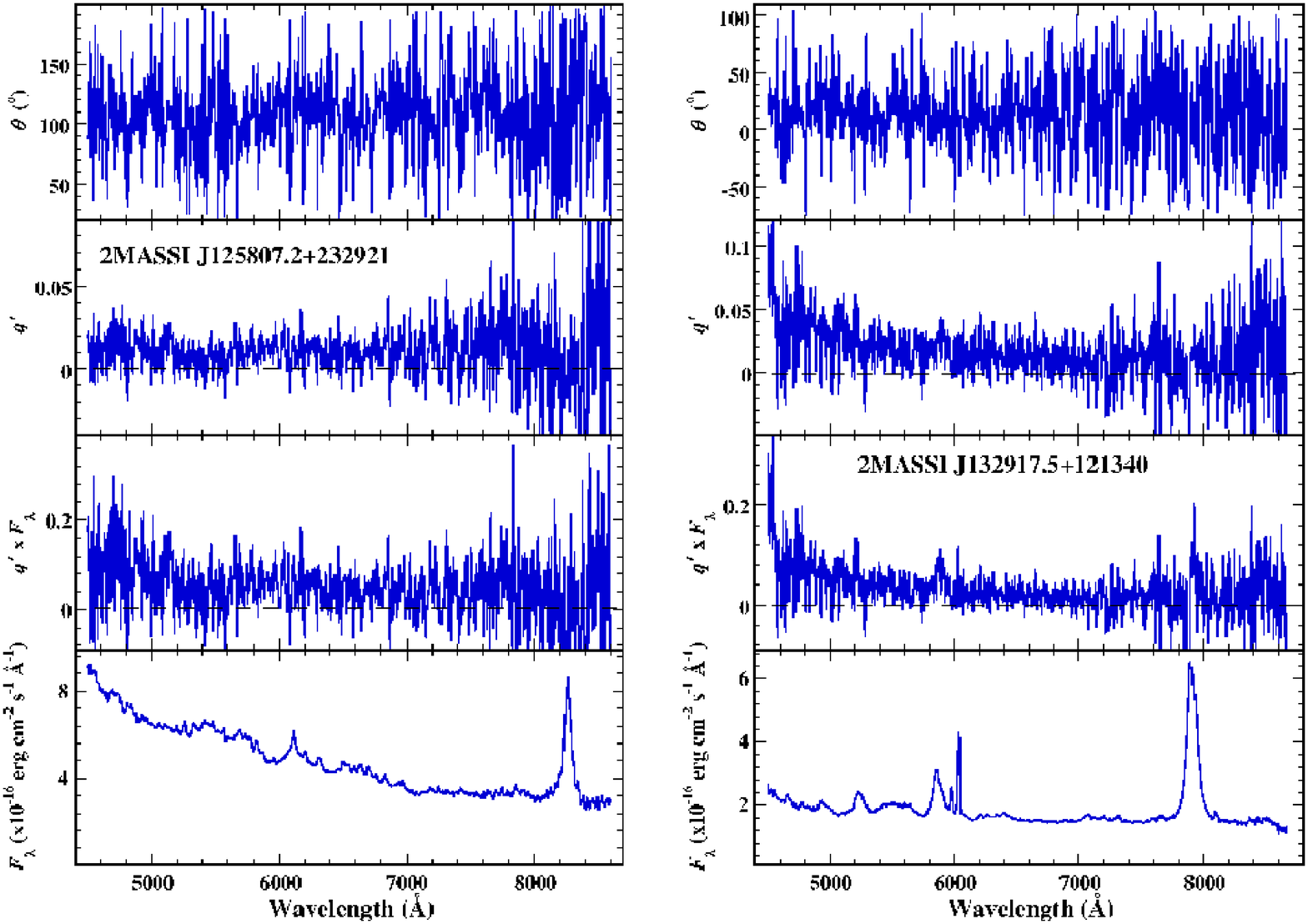}
\caption{Spectropolarimetry of Type~1 2MASS QSOs.  For each object the 
panels display (from bottom to top) the spectra of total flux density
($F_\lambda$),
Stokes flux density ($q' \times F_\lambda$) for a frame aligned with 
the mean polarization position angle at 5000--8000~\AA ,
rotated Stokes parameter
($q'\/$), and the equatorial polarization position angle ($\theta\/$).
\label{fig1}}
\end{figure}
\clearpage

\begin{figure}
\figurenum{1}
\vspace{8.6in}
\includegraphics{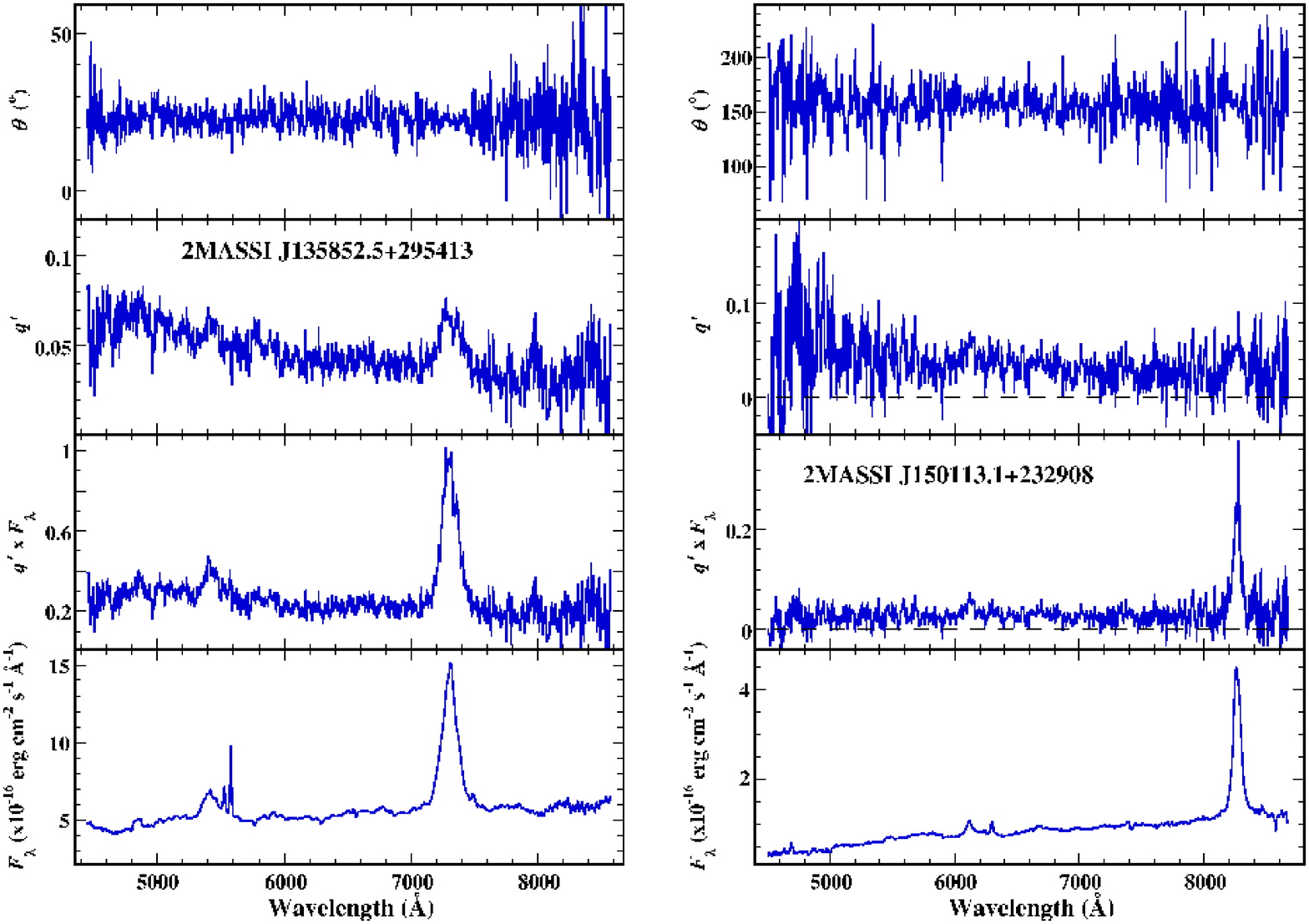}
\includegraphics{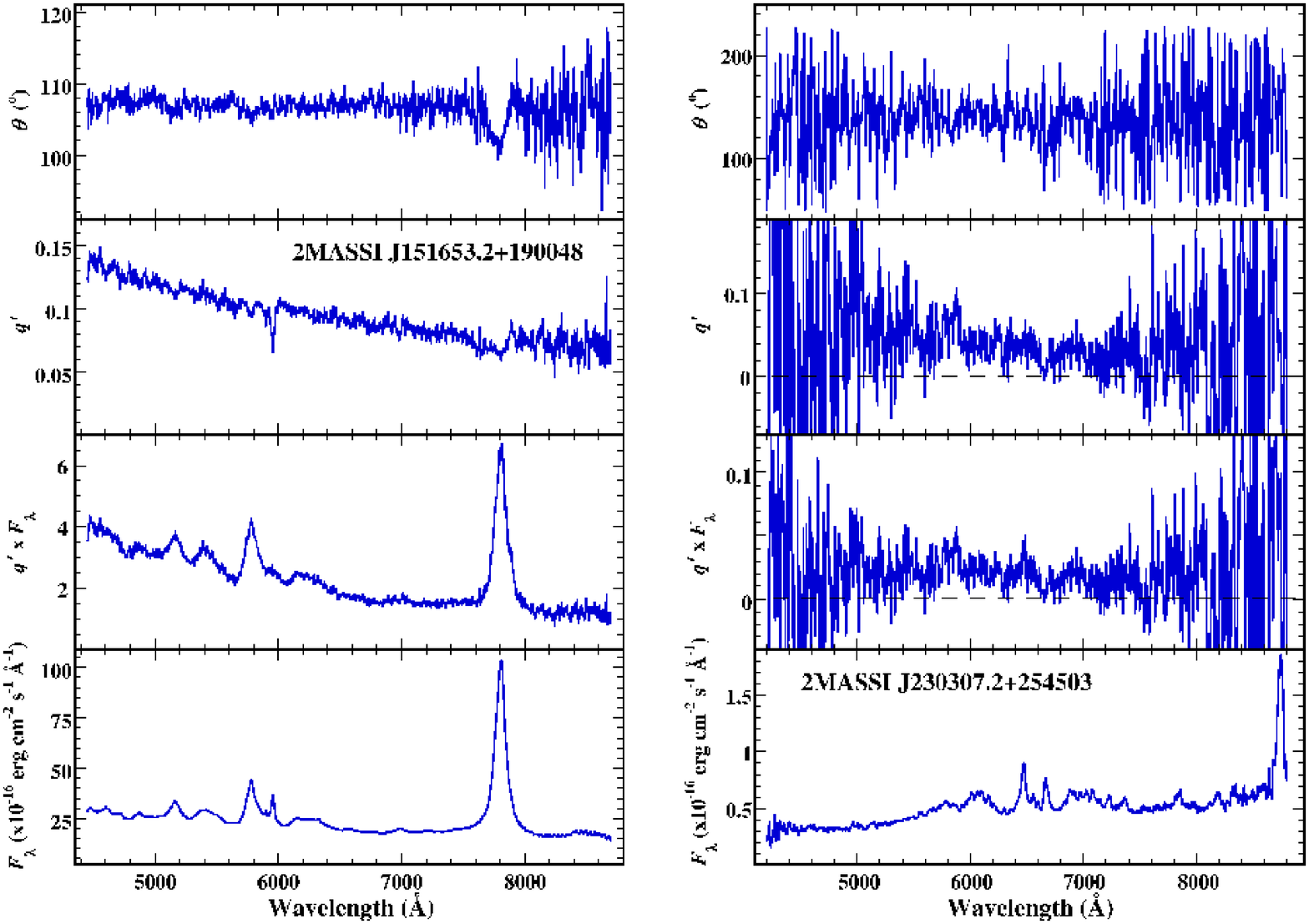}
\caption{Continued.
\label{}}
\end{figure}
\clearpage

\begin{figure}
\figurenum{2}
\vspace{8.6in}
\includegraphics{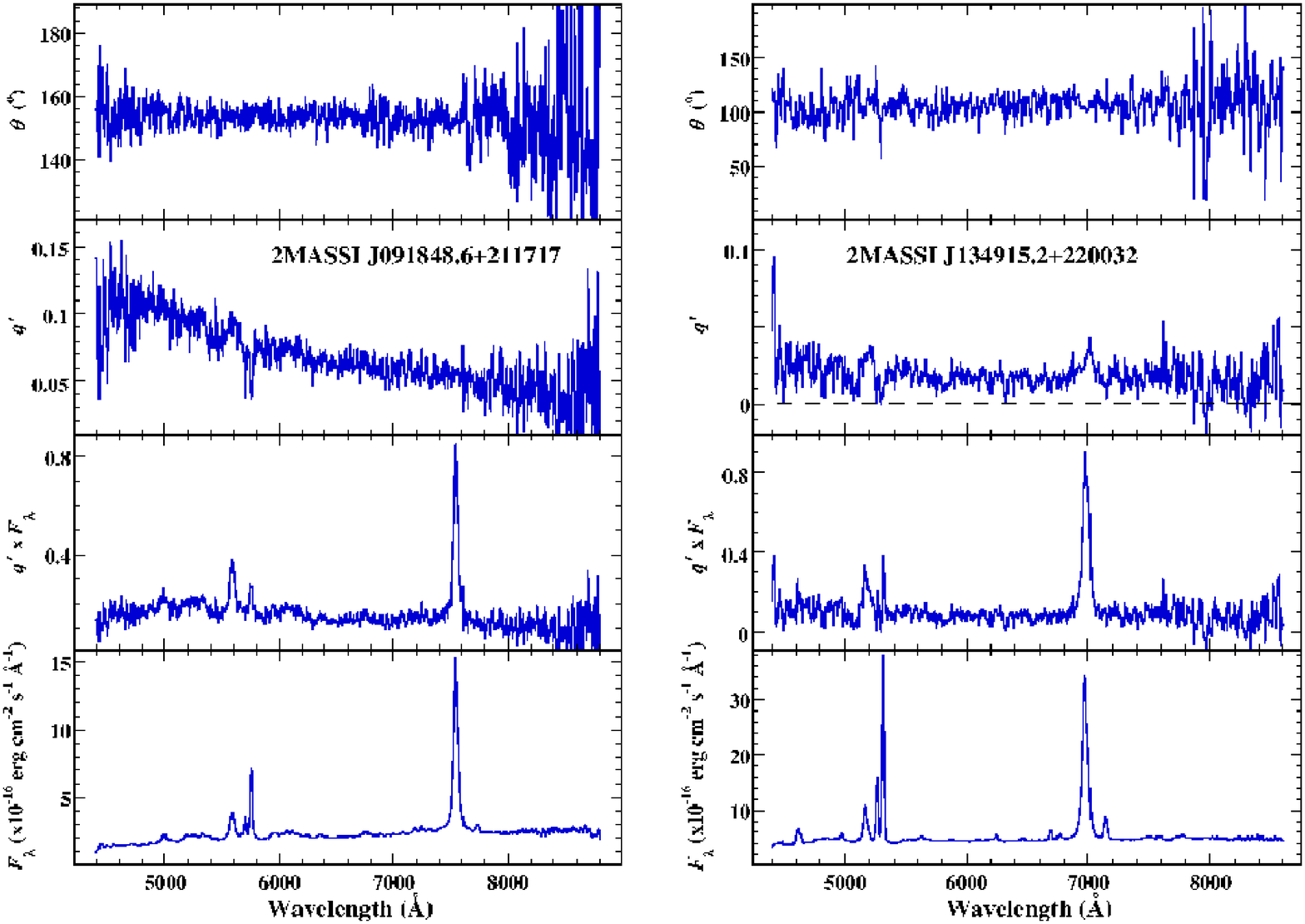}
\includegraphics{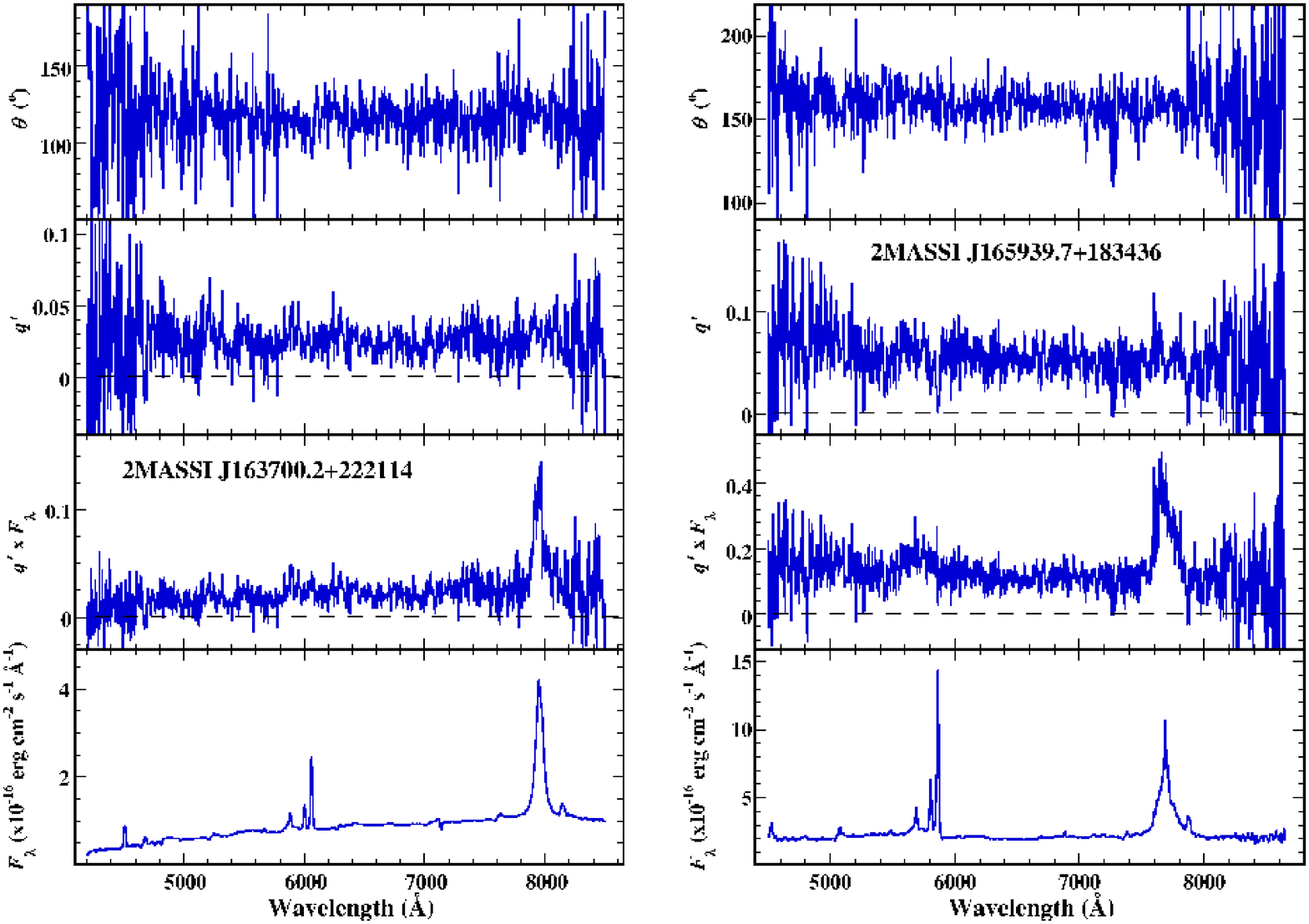}
\caption{Spectropolarimetry of Type 1.5 2MASS QSOs.  The data are
displayed in the same format as in Figure~1.
\label{2_ab}}
\end{figure}
\clearpage

\begin{figure}
\figurenum{2}
\vspace{8.6in}
\includegraphics{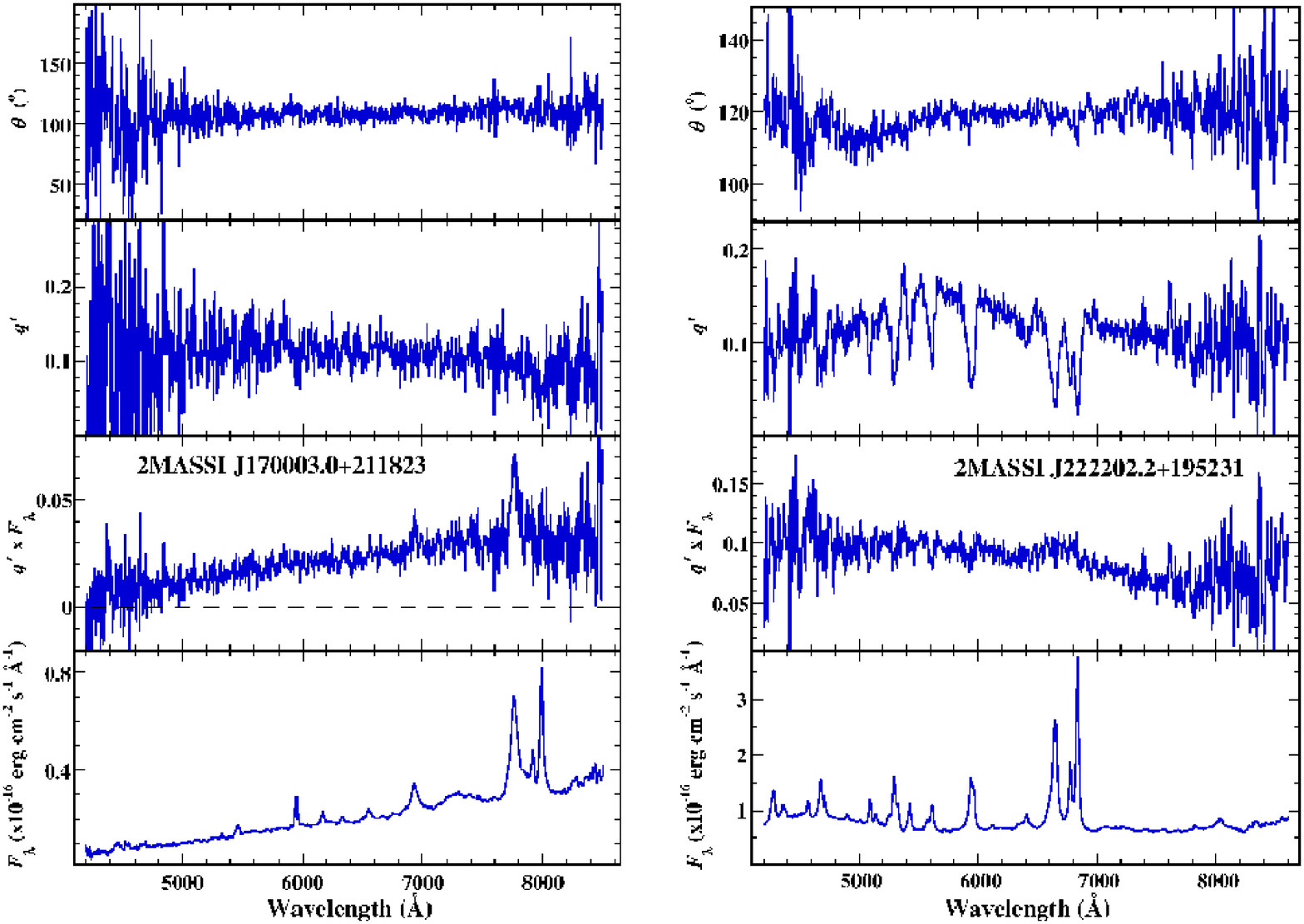}
\includegraphics{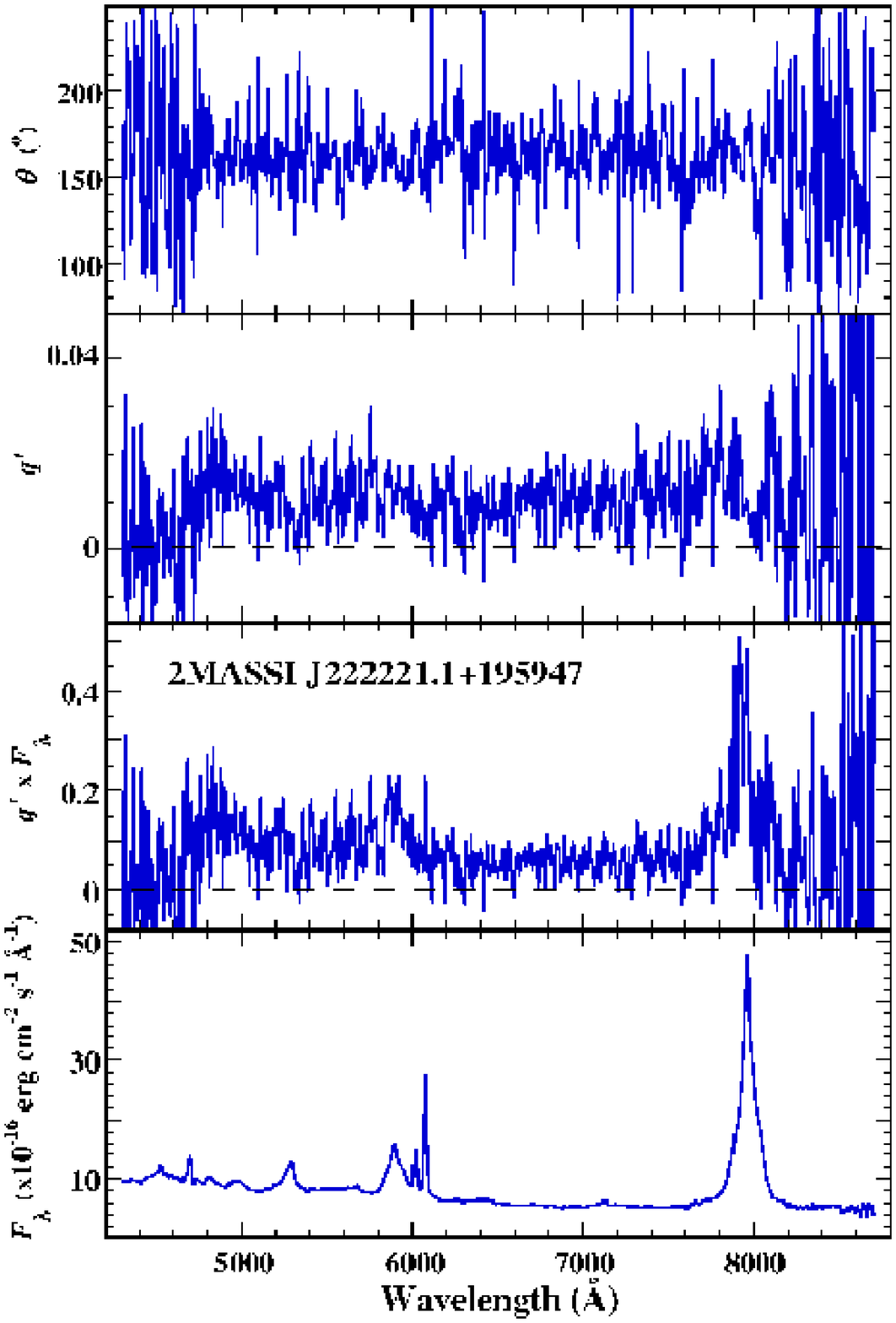}
\caption{Continued.
\label{2_cd}}
\end{figure}
\clearpage

\begin{figure}
\figurenum{3}
\vspace{8.6in}
\includegraphics{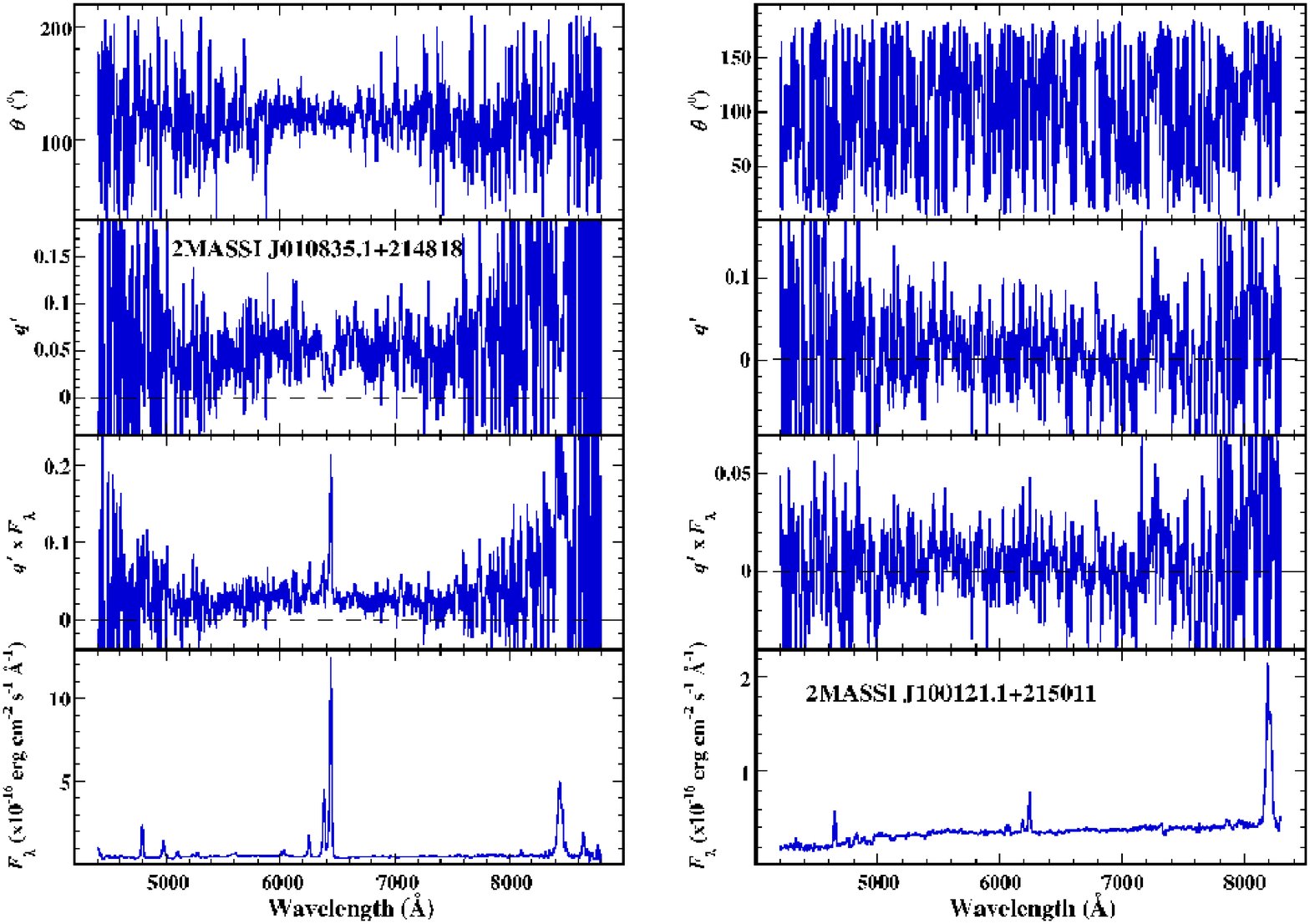}
\includegraphics{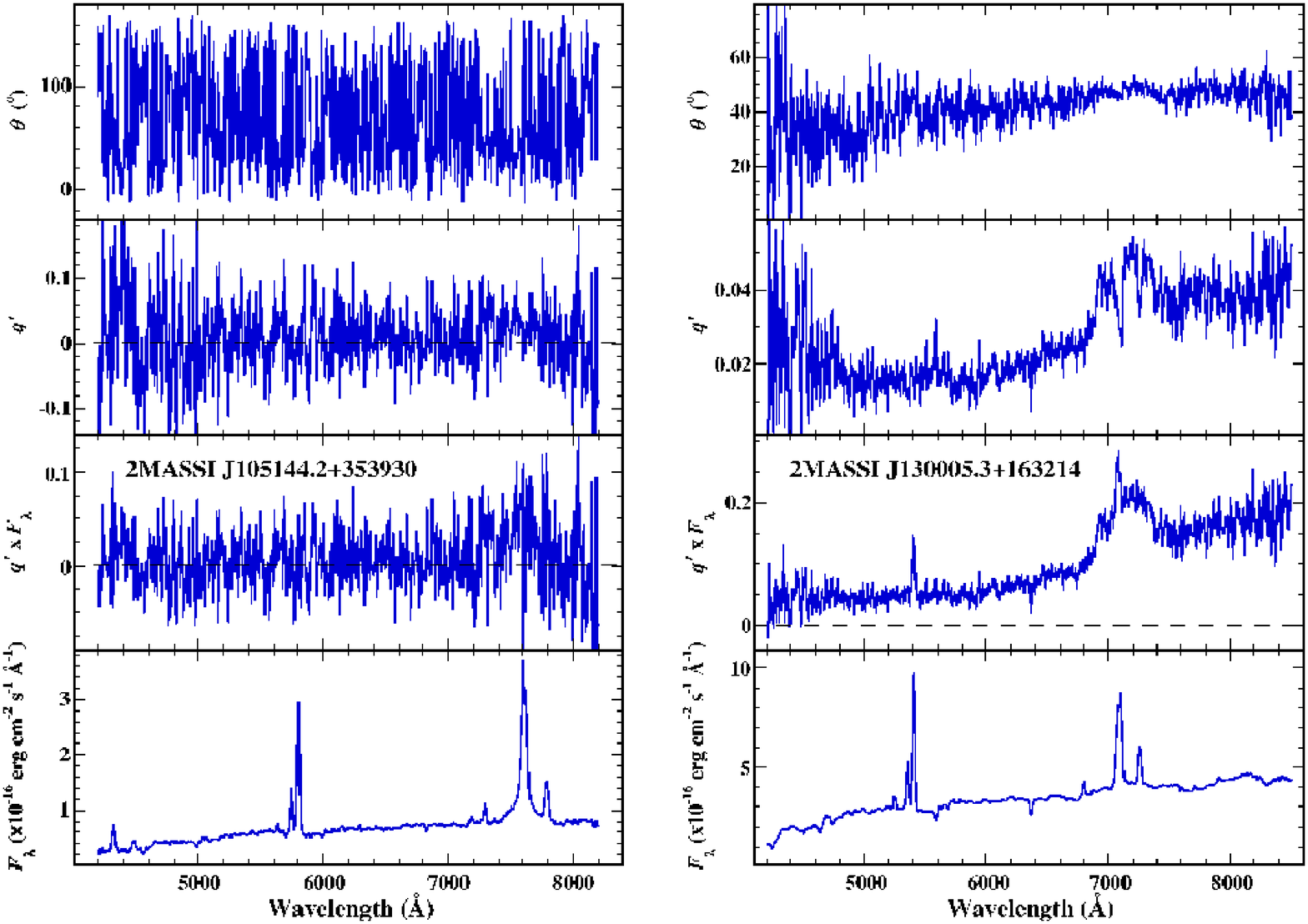}
\caption{Spectropolarimetry 2MASS QSOs of Types 1.8, 1.9, and 2.  The data are
displayed in the same format as in Figure~1.
\label{3}}
\end{figure}
\clearpage

\begin{figure}
\figurenum{3}
\vspace{4.5in}
\includegraphics{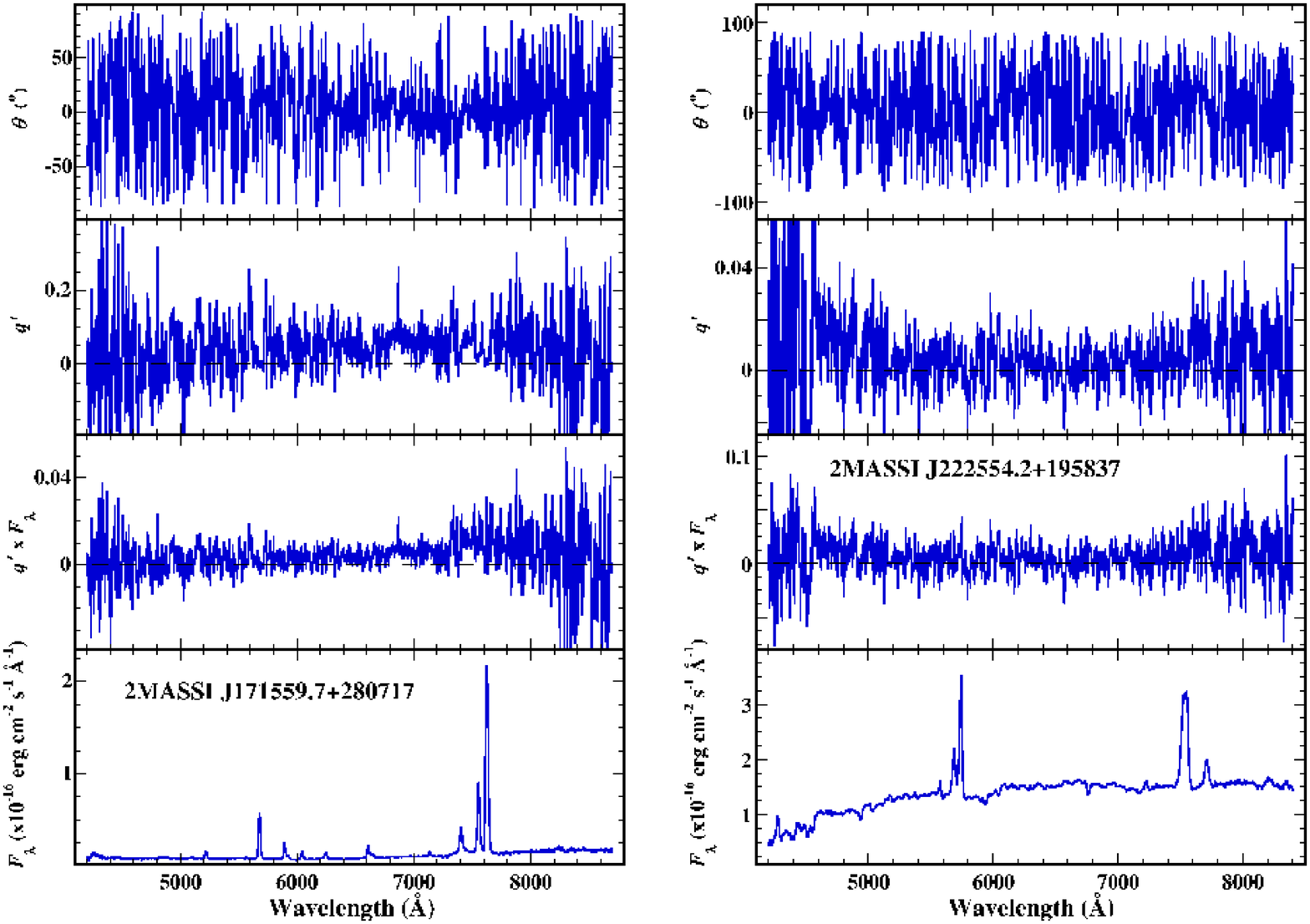}
\caption{Continued.}
\end{figure}
\clearpage

\begin{figure}
\vspace{6.0in}
\figurenum{4}
\includegraphics{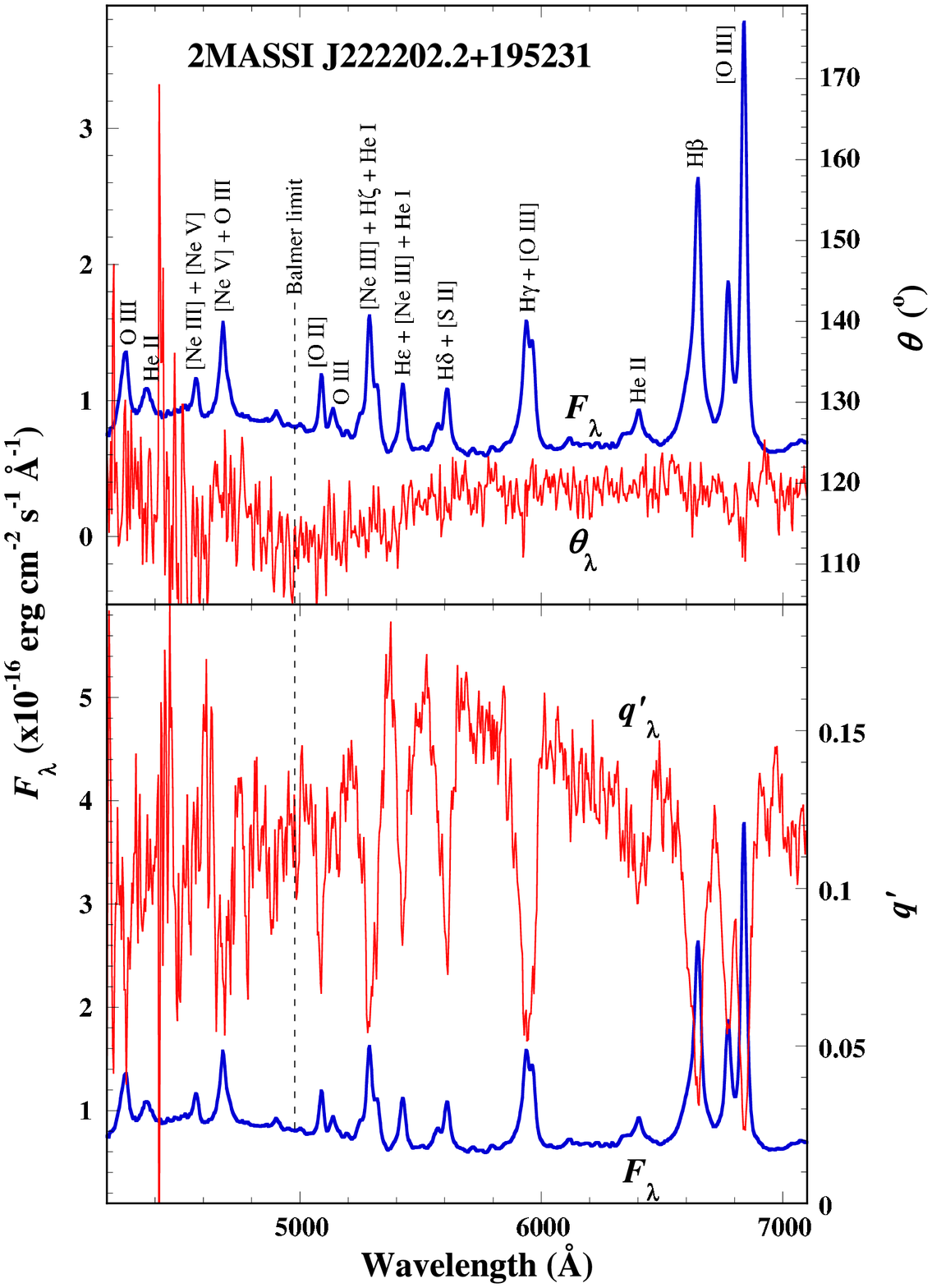}
\caption{A detailed view of the flux spectrum and polarization
of 2M222202.
Major emission lines are identified for the total flux spectrum in
the top panel which compares $F_\lambda\/$ with the spectrum of the 
polarization position angle.
The total flux spectrum is again displayed in the bottom panel for
direct comparison with with the rotated Stokes parameter $q'\/$.} 
\end{figure}
\clearpage

\begin{figure}
\vspace{4.0in}
\figurenum{5}
\includegraphics{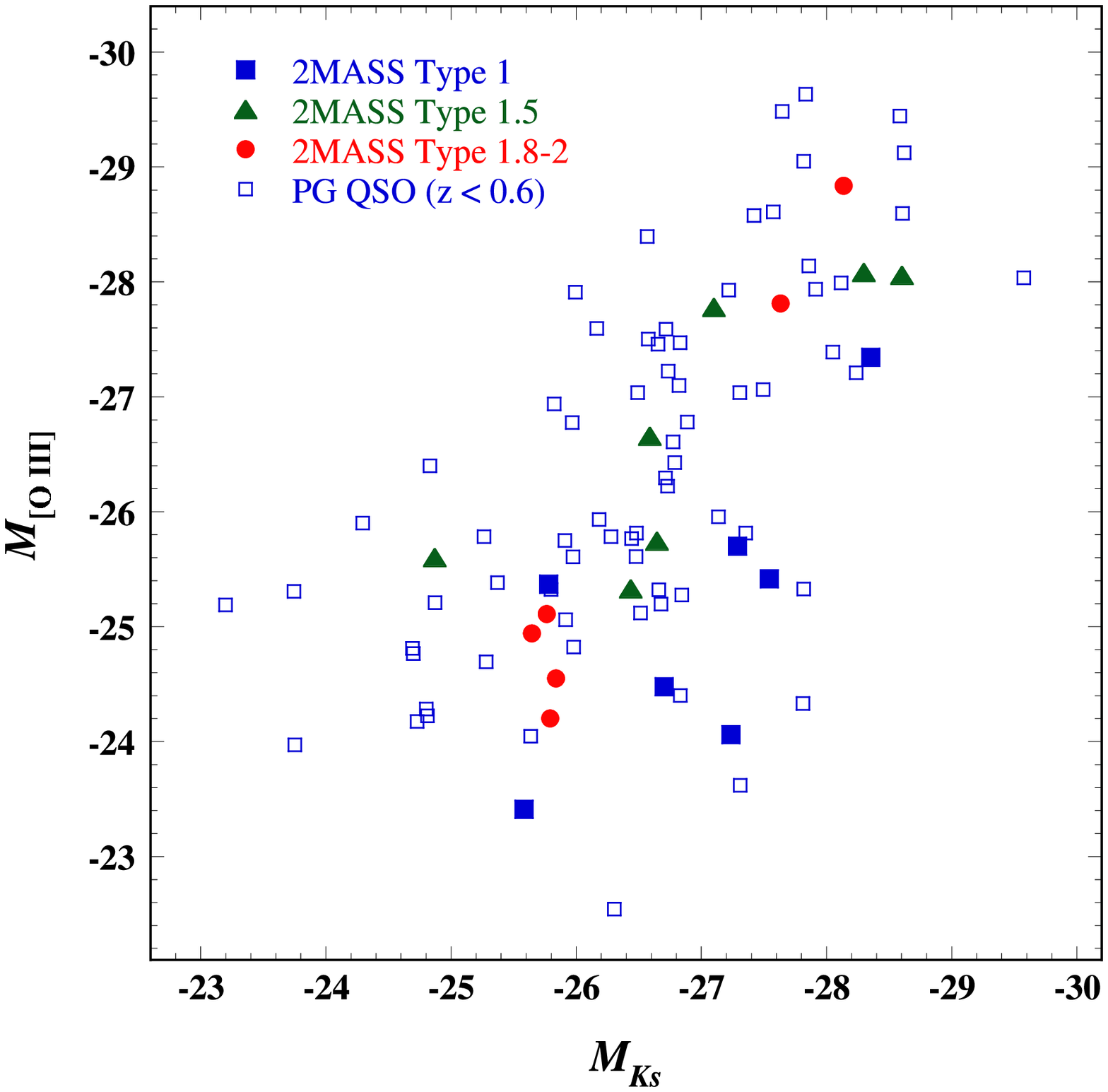}
\caption{Luminosity of [\ion{O}{3}]$\lambda$5007 plotted against
$M_{K_s}\/$ for the 2MASS spectropolarimetry sample and PG QSOs
with $z < 0.6$.
The measure of [\ion{O}{3}] luminosity, $M_{\rm [O III]}\/$, is adopted
from \citet{boroson92} (see text).
The near-IR luminosity of the PG QSOs is determined from the 
observations of \citet{neugebauer87} and assume 
$H_0 = 75$~\kms~Mpc$^{-1}$ and $q_0 = 0$.}
\end{figure}

\begin{figure}
\figurenum{6}
\vspace{3.8in}
\includegraphics{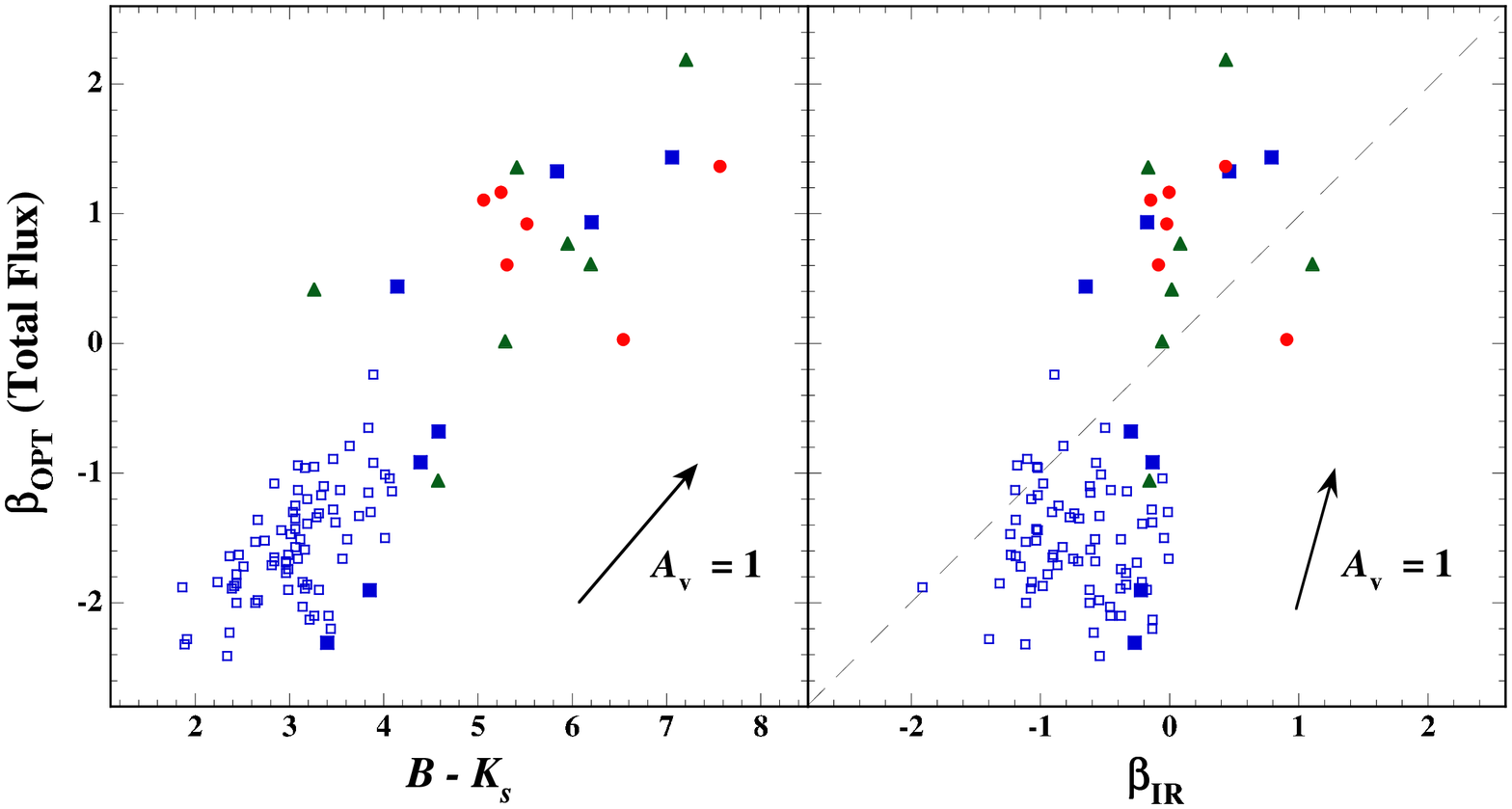}
\caption{{\it Left Panel:\/} The spectral index of the power-law fits
to the optical continua as a function of $B - K_s\/$ color index.
{\it Right Panel:\/} Optical spectral index plotted against the spectral
index of the power-law fits to the near-IR continua. 
The dashed line indicates $\beta_{\rm OPT} = \beta_{\rm IR}$.
Symbols are the same as in Figure~5 for both panels.
The near-IR and $B\/$-band data for the 2MASS QSOs are
from the 2MASS Point Source Catalog.  PG QSO data are from
\citet{neugebauer87}.
For consistency with the PG QSOs data, no correction has been made
for the stellar continuum of the host galaxy.
A reddening vector of $A_V = 1$ is shown in both panels.}
\end{figure}
\clearpage

\begin{figure}
\figurenum{7}
\vspace{4.0in}
\includegraphics{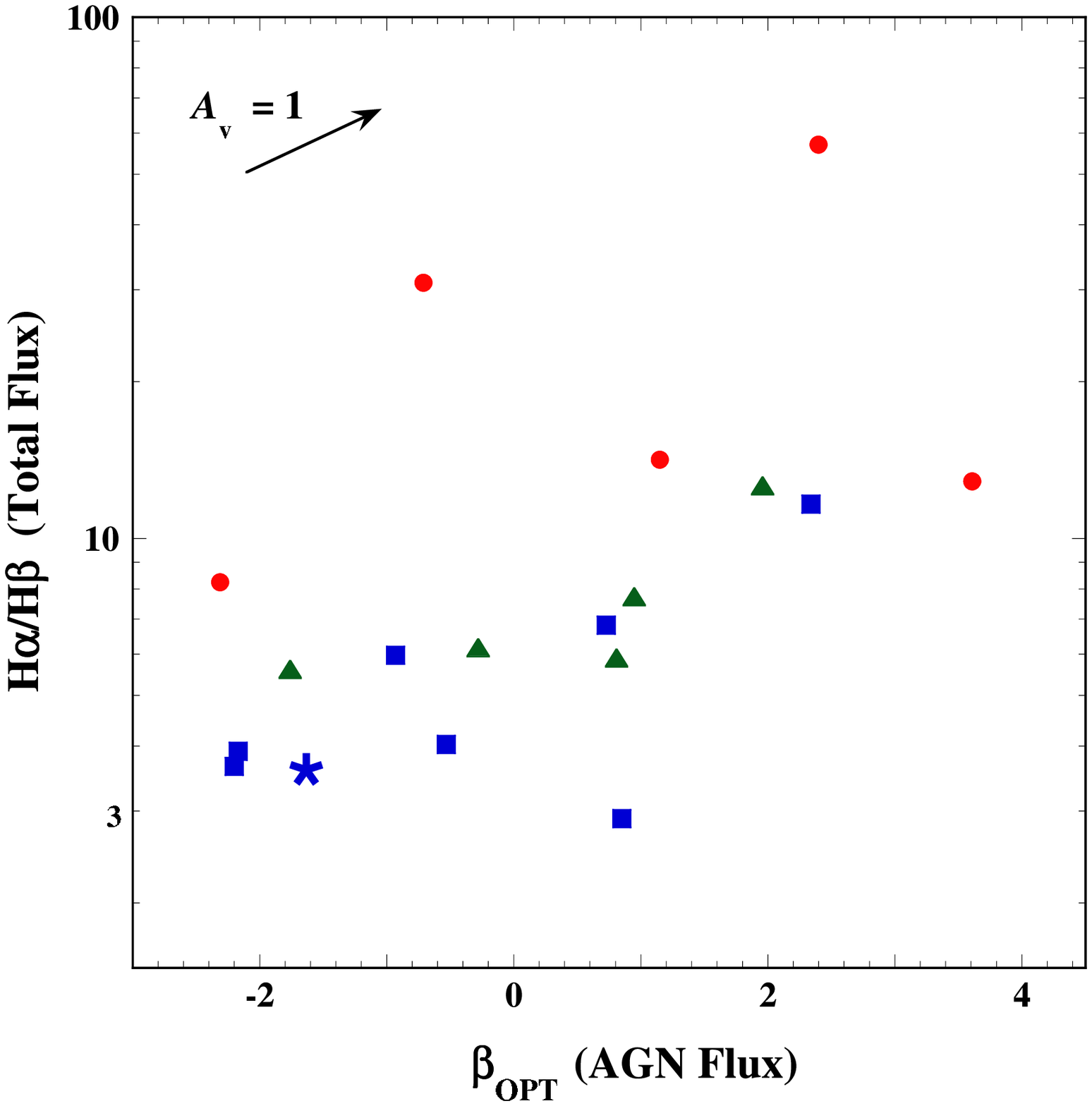}
\caption{The Balmer decrement plotted against $\beta_{\rm OPT}$ for
17 2MASS QSOs in the spectropolarimetry survey.
Starlight from the host galaxies of the 2MASS QSOs has been subtracted
before fitting a power law to the continuum.
It can be seen that in general the Balmer decrement is larger for
objects with redder optical continua.
Symbols are the same as in Figure~5.
The {\it star\/} symbol marks the position of the median $\beta_{\rm OPT}$
for low-redshift PG QSOs and the Balmer decrement of the SDSS
composite QSO.
As in Figure~6, an $A_V = 1$ reddening vector is displayed.}
\end{figure}

\begin{figure}
\figurenum{8}
\vspace{4.0in}
\includegraphics{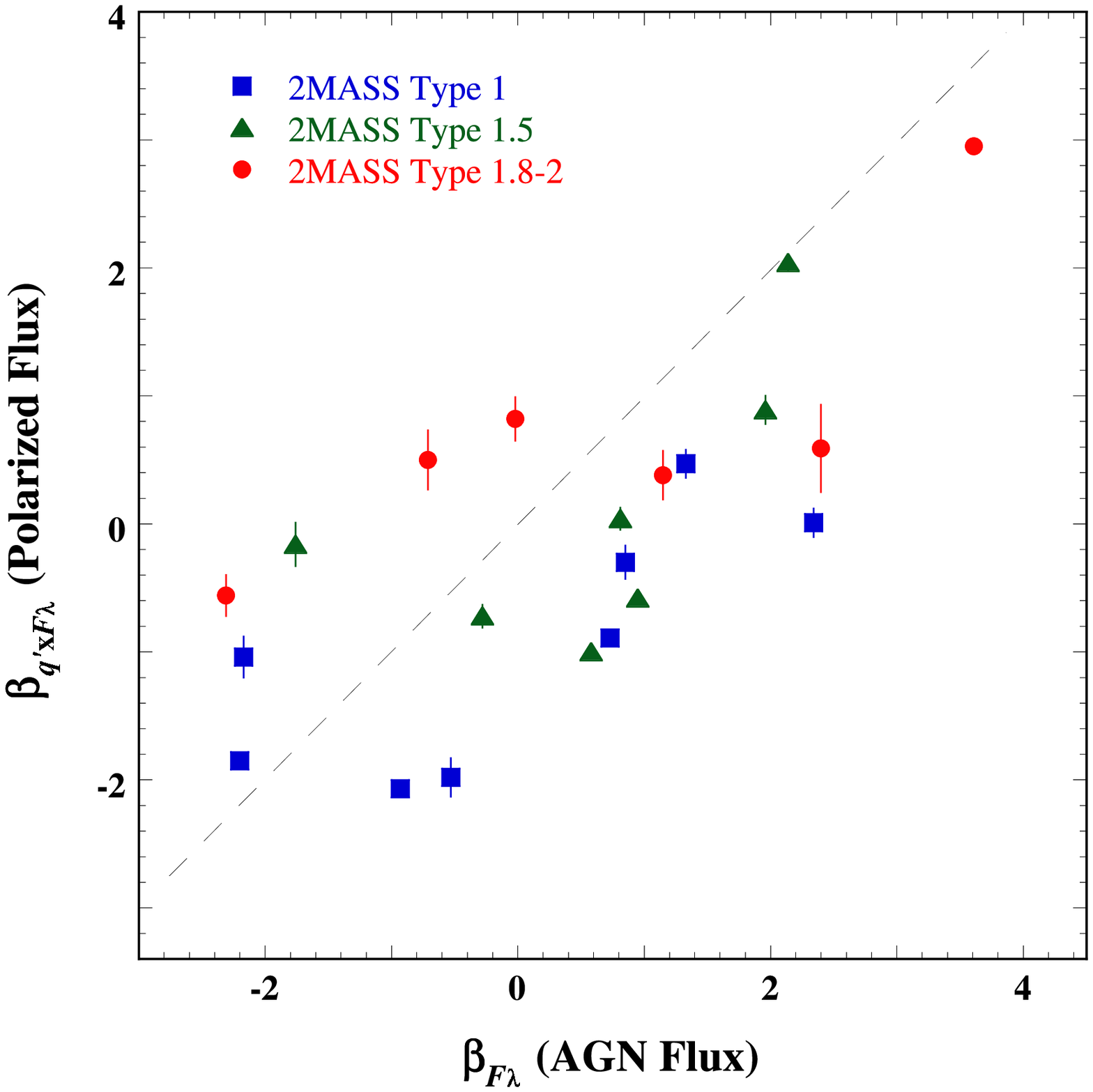}
\caption{The spectral index of the polarized flux density spectrum
plotted against the total flux density spectral index of the AGN component.
The {\it dashed\/} line represents 
$\beta_{q' \times F_\lambda} = \beta_{F_\lambda}$.
Symbols are the same as in Figure~5.
Error bars reflect the uncertainty in the power-law fits to the
spectra, but do not include the uncertainties in the stellar 
spectra subtracted to yield the AGN spectra.
This source of uncertainty is largest for the objects
dominated by the light of the host galaxy; generally, the 
Type~1.8--2 QSOs.}
\end{figure}

\end{document}